\documentclass{aa}  

\usepackage[version=3]{mhchem}
\usepackage{graphicx}
\usepackage{subfigure}
\usepackage{caption}
\usepackage{amsmath}
\usepackage{placeins}

\usepackage{txfonts}
\usepackage{natbib}
\bibpunct{(}{)}{;}{a}{}{,} 

\newcommand{\bi}{\begin{itemize}}
\newcommand{\ei}{\end{itemize}}
\newcommand{\be}{\begin{enumerate}}
\newcommand{\ee}{\end{enumerate}}

\newcommand{\beq}{\begin{equation}}
\newcommand{\eeq}{\end{equation}}
\newcommand{\Hn}{H$_{\rm nuc}$}

\begin{document}


\title{Setting the volatile composition of (exo)planet-building material}
\subtitle{Does chemical evolution in disk midplanes matter?}

\author{Christian Eistrup \inst{1}
\and Catherine Walsh\inst{1}
\and Ewine F. van Dishoeck\inst{1,2}}

\institute{Leiden Observatory, Leiden University, P.O. Box 9513, 2300 RA Leiden, the Netherlands\\
\email{eistrup@strw.leidenuniv.nl, cwalsh@strw.leidenuniv.nl, ewine@strw.leidenuniv.nl}
\and Max-Planck-Institut f\"ur Extraterrestrische Physik, Giessenbachstrasse 1, 85748 Garching, Germany\\}

\date{Received $\cdots$ / Accepted $\cdots$}

\abstract
{The atmospheres of extrasolar planets are thought to be built largely 
through accretion of pebbles and planetesimals. 
Such pebbles are also the building blocks of comets. 
The chemical composition of their volatiles are usually taken to be 
inherited from 
the ices in the collapsing cloud. 
However, chemistry in the protoplanetary disk midplane can modify the 
composition of ices and gases.}
{To investigate if and how chemical evolution affects the abundances and distributions of key 
volatile species in the midplane of a protoplanetary disk in the 0.2--30 AU range.
}
{A disk model used in planet population synthesis models is adopted,
providing temperature, density and ionisation rate at different radial 
distances in the disk midplane.
A full chemical network including gas-phase, gas-grain interactions and 
grain-surface chemistry is used to evolve chemistry 
in time, for 1 Myr. Both molecular (inheritance from the parent cloud) and atomic (chemical reset) initial 
conditions are investigated.}
{Great diversity is observed in the relative abundance ratios of the 
main considered species: 
\ce{H2O}, CO, \ce{CO2}, \ce{CH4}, \ce{O2}, \ce{NH3} and \ce{N2}.  
The choice of ionisation level, the choice of initial abundances, as 
well as the extent of chemical reaction types included 
are all factors that affect the chemical evolution. 
The only exception is the inheritance scenario with a 
low ionisation level, which results in negligible changes 
compared with the initial abundances, regardless of whether grain-surface 
chemistry is included. 
The grain temperature plays an important role, especially in the critical 
20-28 K region where atomic H no longer sticks 
long enough to the surface to react, but atomic O does.  
Above 28 K, efficient grain-surface production of \ce{CO2} ice is seen, 
as well as \ce{O2} gas and ice
under certain conditions, at the expense of H$_2$O and CO. 
\ce{H2O} ice 
is produced on grain surfaces only below 28~K. For high ionisation levels at
intermediate disk radii, 
\ce{CH4} gas is destroyed and converted into CO and \ce{CO2} (in contrast with previous models), 
and similarly \ce{NH3} gas is converted into N$_2$. At large radii around 30 AU, \ce{CH4} ice is enhanced leading to a low gaseous CO abundance. As a result, the overall C/O ratios for gas and ice
change significantly with radius and with model assumptions. For high ionisation levels, chemical processing becomes significant after a few times $10^{5}$~yrs.}
{Chemistry in the disk midplane needs to be considered in the determination 
of the volatile composition of planetesimals.  
In the inner $<$30 AU disk, interstellar ice abundances are preserved only 
if the ionisation level is low, 
or if these species are included in larger bodies within $10^{5}$~yrs.}

\keywords{protoplanetary disks -- planet formation -- astrochemistry -- planetary atmospheres -- molecular processes}

\maketitle


\section{Introduction}
\label{intr}

The discovery of more than 2000 extrasolar planets by the radial velocity
and transiting techniques
\citep[e.g.,][]{udry2007,borucki2011,batalha2013,fischer2014} 
has signaled the next phase in exoplanet research: the
characterization of their atmospheres. Simple molecules such as CO,
\ce{H2O} and perhaps \ce{CO2} and \ce{CH4} are being detected in a
growing number of exoplanet atmospheres
\citep[e.g.,][]{seager2010,snellen2010,birkby2013,fraine2014,crossfield2015,sing2016}.
These atmospheres are thought to be built up largely by the accretion
of pebbles and planetesimals in the natal protoplanetary disk 
(see \citealt{johansen2014} and \citealt{benz2014} for reviews), 
hence the atmospheres should reflect the chemical composition of the disk. 
There are two different views on how to treat the chemistry in the midplanes of disks, 
depending on the scientific focus and heritage.

Planet formation and population synthesis models
\citep[e.g.,][]{ida2004,ida2008,alibert13} consider multiple
physical effects taking place in a protoplanetary disk, such as
gravitational interactions between bodies, orbital excitation and
eccentricity damping, gas drag, accretion of material onto planets,
and planet migration in the gaseous disk. 
Hence, there is a high degree of physical complexity and detail to 
planet formation processes in these simulations. 
However, these models do not contain any detailed chemistry.  
Either they simply use the observed chemical
abundances in interstellar ices and assume that these abundances are preserved
during disk evolution, or they assume that thermodynamic equilibrium
is attained so that chemical abundances are controlled by temperature
and pressure only \citep[e.g.,][]{mousis2010,johnson2012,
moses2013,marboeuf14,thiabaud2015gascomp}. 
The main observational test is through statistical comparisons with the
observed populations of exoplanets and their predicted compositions.

The alternative view starts from detailed physico-chemical models of
protoplanetary disks which are closely linked to, and tested by, a
wide variety of astronomical observations 
(see reviews by \citealt{williams11} and \citealt{armitage2011}).  
Starting from an assumed (static)
surface density distribution, scale height and disk flaring, such
models first determine the temperature structure of dust and gas
heated by the central star through calculation of the full radiative
transfer of the dust and the thermal balance of the gas
\citep[e.g.,][]{dullemond2007,nomura2005,woitke2009,bruderer2013}.
This physical model is then coupled with an extensive gas-grain
chemistry network to solve the kinetic chemistry equations at each
point in the disk and compute the chemical composition of the gas and
ice as a function of time
\citep[e.g.,][]{bergin2007,furuya2014,cleeves2014water,reboussin2015,walsh2012,walsh15}.
Since planets are formed in the midplanes of disks, it is particularly
important to consider the composition and evolution in the midplanes. To what extent
is the initial chemical composition of material that is accreted onto
a protoplanetary disk preserved, and what happens to the material
after it reaches the midplane of the disk, i.e., to what extent is it
reset \citep{visser2009,pontoppidan2014}? Does planetesimal formation
happen so fast that ices are incorporated into large bodies early on
in the evolution, preventing further chemical processing
\citep{zhang2015}?

Additional clues to the chemical evolution in disks come from the
observations of comets in our own solar system \citep{charnley11}.
Cometary records suggest that the chemical composition of the pre-solar
nebula has been at least partially preserved in the comet-forming zone
throughout its lifetime, pointing to little or no
chemical processing.  However, the original composition of the
material that was present in the protoplanetary disk around the Sun
when it formed remains unknown, and studies of other disks are needed
to provide a framework for our own solar system. Particularly
interesting are the recent results from the ESA {\it Rosetta} mission
finding significant amounts of \ce{O2} in comet 67P/C-G
\citep[see][]{bieler15}, with similarly high O$_2$ abundances inferred
for comet Halley from a re-analysis of the {\it Giotto} data
\citep{rubin2015}. Abundances as high as a few~\% of solid \ce{O2} with
respect to solid \ce{H2O} are not yet fully understood. Lastly, the deuteration of water and organics also provides insight into the history of the pristine material from the ISM \citep[see][]{ceccarelli2014}.

\begin{figure*}
\subfigure[][]{\includegraphics[width=0.5\textwidth]{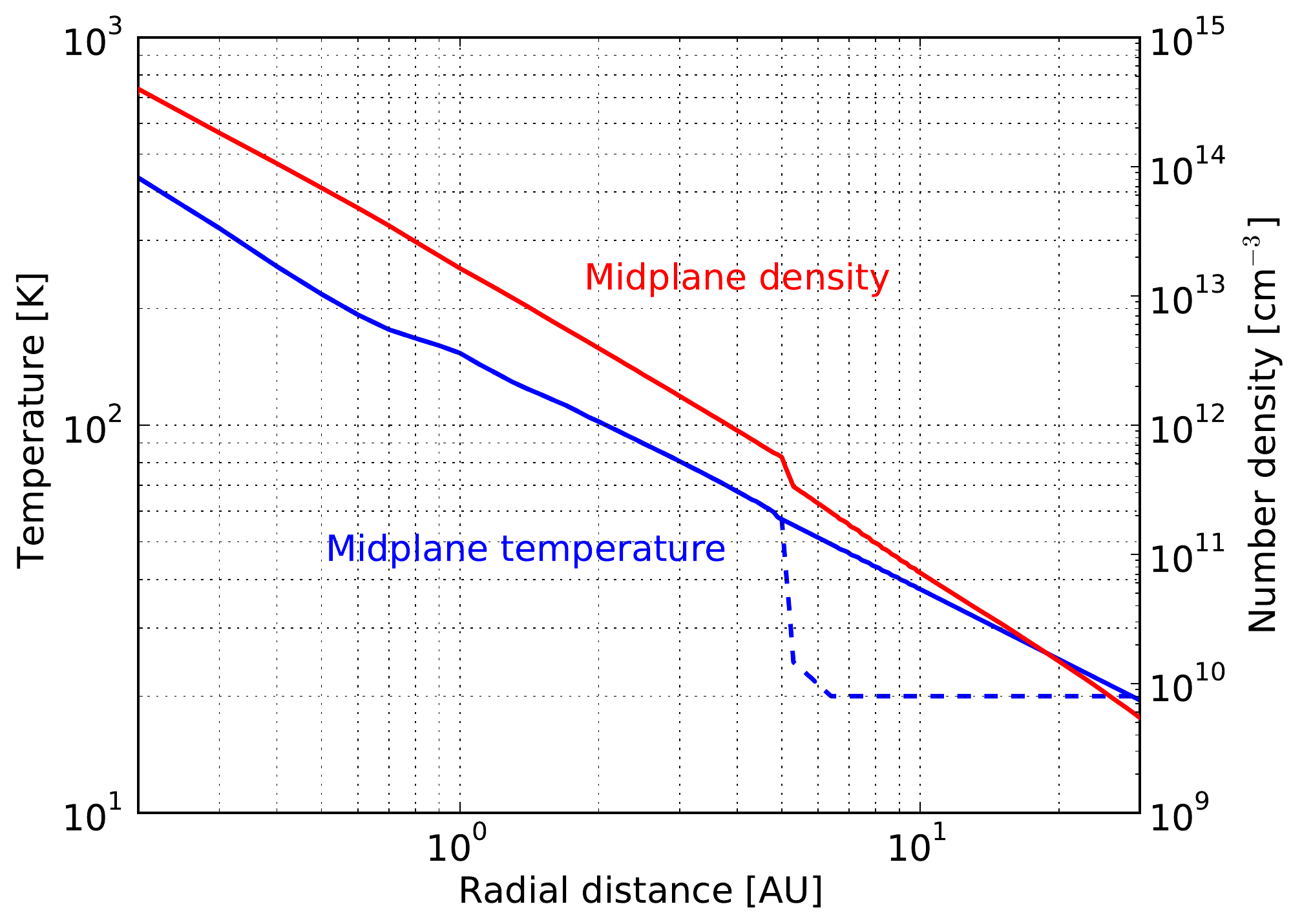}\label{temp_dens}}
\subfigure[][]{\includegraphics[width=0.5\textwidth]{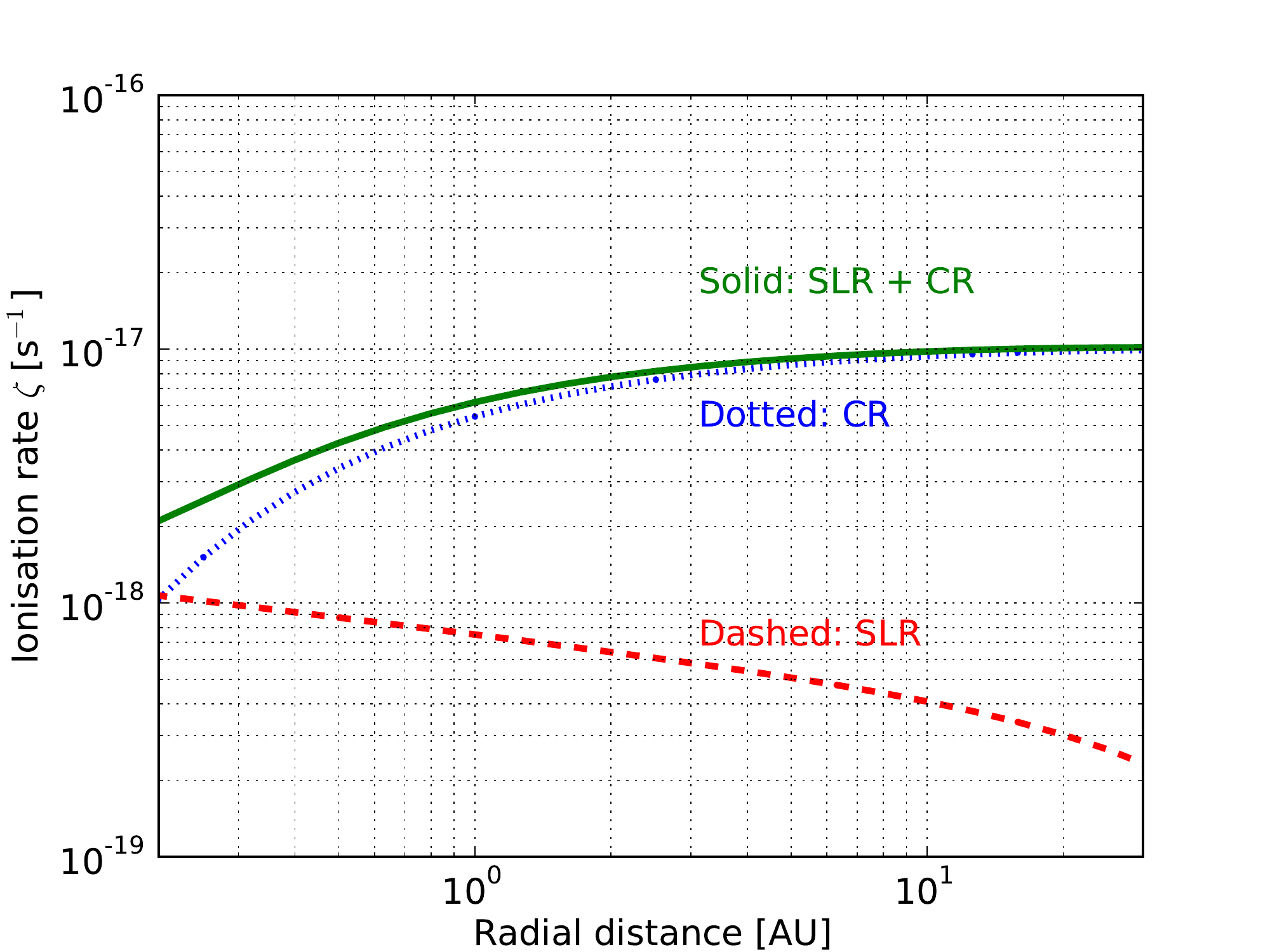}\label{cr_level}}
\caption{(a) The temperature $T(R)$ in K (blue) and number density $n(R)$ in cm$^{-3}$ (red) profiles 
for the disk midplane. 
The solid blue line indicates the adjusted temperature profile (as described in the text). 
The original temperature profile from \citet{alibert13} beyond 5.2~AU is indicated 
by the dashed blue line. 
(b) The ionisation rate $\zeta(R)$ in s$^{-1}$ adopted for the disk midplane. 
The red dashed line depicts the contribution to the ionisation rate from 
short-lived radionuclides (SLRs) only. 
The blue dotted line is the contribution from external cosmic rays (CRs) only. 
The solid green line represents the total ionisation rate (SLRs and CRs) 
as a function of radial distance.}
\label{simple_c}
\end{figure*}

In their planet population synthesis models, \citet{marboeuf14}
assumed the initial chemical abundances to be inherited directly from
the interstellar ices observed in dense interstellar clouds.  A set of
eight volatile molecules (\ce{H2O}, CO, \ce{CO2}, \ce{CH4}, \ce{H2S},
\ce{CH3OH}, \ce{N2}, and \ce{NH3}, species also considered in this
work) were homogeneously distributed in their model disk midplane with
relative ratios consistent with interstellar ice observations
\citep{gibb2004,oberg2011ices,boogert2015}. Depending on the physical
conditions in different parts of the midplane, as well as the
sublimation temperatures of the species, these molecules could then
either be assigned to the gas or ice, with the threshold set by the
icelines of the species. Icelines (or snowlines) mark the radius in
the disk midplane beyond which species exist solely in ice form and
are thus depleted from the gas.  This occurs at the radius where the
accretion rate onto grain surfaces (or freezeout) exceeds the
desorption rate from grain surfaces due to the negative temperature
gradient in the midplane.  The relative rates of these processes are
very strong functions of temperature leading to a narrow transition
region from gas to ice (moving outwards in radius).  The position of
the midplane iceline for a particular species will depend on its
volatility (i.e., its binding energy).  \citet{marboeuf14} do not
consider any chemical reactions in their models, besides freezeout and
desorption.

The positions of the icelines are important because they determine
which species are gas and ice at any location in the disk, and thus
which material is available to build larger bodies (solids only). If, for example, a
giant planet is forming in the disk, the composition of its core will
reflect the ice compositions at the different positions in the disk
through which the forming planet has moved. The composition of the	
planet's atmosphere, on the other hand, will reflect the gas
composition at the position where the planet becomes massive enough to
accrete an atmosphere onto its surface from the surrounding gas in the
disk. Moreover, accretion of icy bodies may still pollute the
atmosphere. These pebbles and planetesimals migrate through the disk
due to radial drift and may therefore have originated at larger
radii. Depending on the pebble and planetesimal sizes, the migration of these objects also affects the location of the icelines \citep[see, e.g.,][]{piso2015}.

Particularly important is the C/O ratio of the solid and gaseous
material in the disk \citep{oberg2011co}. The ratio depends not only on the
different volatilities of the chemical species but also on their production or
destruction as a consequence of chemical processing.  Since \ce{H2O}
and \ce{CO2} (which are both O rich) freeze out at higher temperatures
than species that are more C rich, such as \ce{CH4} and CO, the C/O
ratio depends on both the physical structure and chemistry in the
disk. Ultimately, the chemical composition of a planet's core and its
atmosphere may thus differ depending on the history of the disk, the
formation location of the planet, and any subsequent migration.

To address these questions, we use a physical disk model, in
particular its midplane temperature and density, which is the same as
that considered in the \citet{marboeuf14} population synthesis models.
We compute the abundances of chemical species with time using a
comprehensive chemical network and different sets of assumptions (see
below) to investigate the degree to which chemical
evolution/processing affects the resulting abundances of key
volatiles in the disk midplane. The sensitivity of our results to
the choice of (i) initial chemical abundances (parent cloud inheritance or chemical reset), 
(ii) the physical conditions (in particular ionisation level), and
(iii) the types of chemical reactions included in the model, are also investigated, 
with details provided in Sect.~\ref{methods}.  
This generates eight different simulations, the results of which are presented
in Sect.~\ref{results}. 
Sect.~\ref{discussion} discusses the validity of the 
inheritance and reset scenarios, the implications for planet 
formation, and the extent to which the results hold for other 
disk models.   
Sect.~\ref{conclusions} summarises the conclusions from this work.

\begin{table}
\caption{Initial abundances (with respect to H$_{\rm nuc}$) for atomic and molecular initial abundances setups. 
The binding energies $E_{b}$ for all species are also listed.}             
\centering                          
\begin{tabular}{l l c r}              
\hline\hline                        
Species 	& Atomic 		      & Molecular     &$E_{b}$[K]\\              
\hline                              
\\
   H		& 9.1$\times 10^{-5}$ & 5.0$\times 10^{-5}$    & 600\\      
   He	 	& 9.8$\times 10^{-2}$ & 9.8$\times 10^{-2}$    & 100\\
   \ce{H2} 	& 5.0$\times 10^{-1}$ & 5.0$\times 10^{-1}$    & 430\\
   N 		& 6.2$\times 10^{-5}$ &                                    & 800\\
   O 		& 5.2$\times 10^{-4}$ &                                    & 800\\ 
   C		& 1.8$\times 10^{-4}$ &                                    & 800\\
   S		& 6.0$\times 10^{-6}$ &                                    & 1100\\
   \ce{H2O}	&		      &3.0$\times 10^{-4}$         & 5770\\
   CO		        &		      &6.0$\times 10^{-5}$         & 855\\
   \ce{CO2}	&		      &6.0$\times 10^{-5}$         & 2990\\
   \ce{CH4}	&		      &1.8$\times 10^{-5}$         & 1090\\
   \ce{N2}	&		      	      &2.1$\times 10^{-5}$         & 790\\
   \ce{NH3}	&		      &2.1$\times 10^{-5}$         & 3130\\
   \ce{CH3OH}	&		      &4.5$\times 10^{-5}$         & 4930\\
   \ce{H2S}	&		      &6.0$\times 10^{-6}$         & 2743\\
   \ce{O2}	&		      	      &0                                      & 1000\\
   \ce{HCN}	&		      &0                                      & 3610\\
   \ce{NO}          &		      &0                                      & 1600
   \\
   \hline                                   
\label{init_abun}
\end{tabular}
\end{table}

\section{Methods}
\label{methods}

\subsection{Physical disk model}
\label{phys_setup}

The protoplanetary disk is taken from the models of
\citet{alibert13}, \citet{marboeuf14}, and \citet{thiabaud2015gascomp} which provide the midplane temperature
$T(R)$, pressure $p(R)$, and surface density profiles $\Sigma(R)$ with radius, $R$.
The disk has a parameterised surface density profile \footnote{\citealp[From][Eq. (1)]{marboeuf14}},  
\beq
\Sigma(R)=\Sigma_{0}\cdot\left(\frac{R}{5.2\mathrm{AU}}\right)^{-\gamma}
\cdot\mathrm{exp}\left(\frac{-R}{a_{C}}\right)^{2-\gamma}, 
\label{eq_sigma}
\eeq 
where, $a_{C} = 20$~AU, $\gamma = 0.8$ \citep[see the prescription in][Table 1]{alibert13}, 
and $a_{C}$ and $\gamma$ are constrained by observations \citep{andrews10}. 
The surface density is $\Sigma(5.2$ AU$)=16$~g cm$^{-2}$ at $R =5.2$ AU and   
the disk is truncated at $R_{\rm out} = 30$~AU. The total mass of the disk is 

\begin{align*}
M_{\mathrm{disk}} &=\int_{R_{0}}^{R_{max}} 2\pi R \Sigma(R)\,dR\ = 1.3\times 10^{-3}M_{\odot}\approx 0.13 \mathrm{MMSN},
\end{align*}

with $R_{0}=0.05$ AU, $R_{max}$ defined as the radius at which the cumulative disk mass calculated from inside out reaches the total disk mass within 1\%, $\Sigma(R)$ taken from Eq. \ref{eq_sigma}, and MMSN = $1\times 10^{-2}M_{\odot}$ from \citet{weidenschilling1977}.

The focus here is on the midplane of the disk, where the gas and dust temperatures are assumed to be coupled. Physical disk models have been developed to explain a wide variety of observations where the emission usually arises from higher up in the disk atmosphere, and where gas/grain decoupling for temperature is significant. However, since the disk vertical structure is not relevant for planet formation in the midplane, a midplane-only model is used here.

The radial grid used here consists of 119 points from $R = 0.2$~AU to $R = 30$~AU, 
with radial step sizes of $\Delta R=$ 0.1 AU and 1 AU, 
inside and outside of 10 AU, respectively.

\begin{table*}[]
\centering   
\caption{Chemical reaction types included in the two versions of the chemical network.}             
\begin{tabular}{l c c}        
\hline\hline                 
Reaction type & Reduced chemical network & Full chemical network \\    
\hline                        
   Two-body gas-phase reactions 	& x & x \\      
   Direct cosmic ray ionisation 	& x & x \\
   Cosmic ray-induced photoreaction 	& x & x \\
   Grain-cation recombination 		& x & x \\
   Freezeout				& x & x \\
   Thermal desorption 			& x & x \\
   \hline
   Photodesorption 					& & x \\
   Grain-surface cosmic ray-induced photoreaction	& & x\\
   Grain-surface two-body reaction 			& & x \\
\hline                                   
\label{reac_types}
\end{tabular}
\end{table*}

The disk model includes irradiation from a central star with
a spectral type similar to the Sun.  The temperature profile from
\citet{alibert13} is slightly adjusted to remove two features: (i) a
physically unrealistic drop at 5.2~AU, and (ii) an imposed lower
temperature limit of 20~K in the outer disk.  The profile used here
follows their profile in the range 0.2 AU $\leq$ $R$ $<$ 5.3 AU, and
uses a power-law function $T\propto R^{-0.6}$ to extrapolate 5.3 AU
$\leq$ $R$ $\leq$ 30 AU, in agreement with full 2D radiative transfer
models of protoplanetary disks \citep[see, e.g.,][]{bruderer14}.  This
adjustment is shown in Fig.~\ref{temp_dens}.  In this work, the solid
blue temperature profile in Fig.~\ref{temp_dens} is used throughout
the disk, whereas the original \citet{alibert13} profile in the outer
disk is given by the blue dashed profile.  No adjustments are made to
the pressure profile from \citet{alibert13}.

The temperature in the model decreases from $T=434$~K at $R=0.2$ AU to
$T=19.5$~K at $R=30$~AU in the outer disk.  The number density over
this radial range spans about 5 orders of magnitude, reaching almost
$n = 10^{15}$~cm$^{-3}$ close to the star, and dropping to about 
$n = 10^{10}$~cm$^{-3}$ at 30 AU.

Two different levels for the ionisation rate throughout the disk are considered,
a low level and a high level. 
In particular, recent models have shown that cosmic rays can be excluded from 
disks by the combined effects of stellar winds and magnetic field structures 
\citep{cleeves13crex,cleeves14crex}. The purpose of these two levels of ionisation
is to investigate the effect on chemical reactions that are driven by
ionisation.
Fig. \ref{cr_level} shows the ionisation level profiles used in the simulations. 

The case with \emph{low ionisation level} considers ionisation originating from 
the decay products of short-lived radionuclides (henceforth referred to as SLRs) only. 
Low ionisation is labeled as ``SLR''. 
The implementation of this ionisation source into the simulations is done 
using a simplified version of the prescription given in Eq.~30 in 
\citet{cleeves13slr},
\beq
\zeta_{\ce{SLR}}(R)=(2.5 \cdot 10^{-19}\,\mathrm{s}^{-1}) 
\left(\frac{1}{2}\right)\left(\frac{\Sigma(R)}{\mathrm{g~cm}^{-2}}\right)^{0.27},
\label{eq_slr}
\eeq
where $\zeta_{\ce{SLR}}$ is the SLR ionisation rate per \ce{H2} molecule in 
s$^{-1}$ and $\Sigma(R)$ is the surface density 
of a disk at a given radius $R$ (see Eq.~\ref{eq_sigma}). 
The simplification ignores the original time dependence of the ionisation rate, 
as given in \citet{cleeves13slr}; however, including the time dependence will 
change the ionisation rate by no more than a factor of 2.
With this low ionisation level, a higher degree of ionisation is obtained in the inner, 
denser disk midplane compared with the outer disk (see the red dashed curve in Fig.~\ref{cr_level}). 
This is because the SLR ionisation rate scales with the midplane number density of SLRs, 
which is assumed to be homogeneous and thus scales with $\Sigma(R)$, see Eq. \ref{eq_sigma}.

On the other hand, the case of \emph{high ionisation level} (green
solid profile in Fig. \ref{cr_level}) considers contributions from the
decay products of SLRs and from cosmic rays (henceforth referred to as
CRs), originating externally to the disk (high ionisation is labeled
as SLR + CR). Such a high ionisation level has been used in many disk
models in the midplane, \citep[see, e.g.,][]{semenov2004}.  These
CRs are able to penetrate the disk to induce UV photons in the
disk midplane via

\beq
\ce{H2 ->[CR] H2+ ->[e^{*}-] H2^{*} -> H2},\\
\label{cr_reactions}
\eeq 
with the resulting photons generated by radiative decay of $\mathrm{H}_2^*$ 
\citep[see][]{prasad1983}.
The CR-ionisation rate contribution, $\zeta_{\ce{CR}}$, is treated by assuming 
the following parameterised prescription:
\beq
\zeta_{\ce{CR}}(R)\approx \zeta_{0}\cdot \mathrm{exp}\left(\frac{-\Sigma(R)}{96\mathrm{~g~cm}^{-2}}\right),
\label{eq_cr}
\eeq
where $\zeta_{\ce{CR}}$ is the CR-ionisation rate per \ce{H2} molecule as function of radius $R$, 
$\zeta_{0}=10^{-17}\mathrm{s}^{-1}$ is the assumed upper limit to the ionisation rate, 
and the exponential term represents an attenuation effect for high surface densities 
\citep[see, e.g.,][]{umebayashi2009}. 
Hence, a higher CR-ionisation rate is reached in the outer disk
midplane than in the inner disk (opposite to the case with SLRs only), as
represented by the blue dotted profile in Fig.~\ref{cr_level}.  
Mutual neutralisation following collisions of ions with grains can lower the
ionisation level, as shown by \cite{willacy1998}. This effect is taken into
account in this work. Deuterium chemistry, however, is not considered here, but will be addressed in a separate paper with overlapping authors \citep{furuya2016}.

\subsection{Chemical model}
\label{chem_model}
A detailed network is used to compute the chemical evolution of 
all species, in which many different reactions and 
pathways are included.
The gas-phase chemistry is from the latest release of the UMIST Database 
for Astrochemistry \citep[see][]{mcelroy13} termed {\sc Rate}12.  
Gas-grain interactions and grain-surface chemistry are included 
\citep[as described in][and references therein]{walsh15}.
The chemistry is solved time-dependently at each radial grid point in
the disk. The chemical evolution is assumed to be isolated at these
grid points with no exchange of material between the grid points during
the evolution. The differences in chemistry at the
different points are therefore dictated only by the differences in
physical conditions $T(R)$, $n(R)$, and $\zeta(R)$. 
Different chemical species have different volatilities, 
and thus different temperatures below which the phase change from gas to ice occurs. 
For each species a binding energy, $E_{b}$~(K), is adopted. 
These values are given in Table~\ref{init_abun}.

\begin{figure*}
\subfigure{\includegraphics[width=0.5\textwidth]{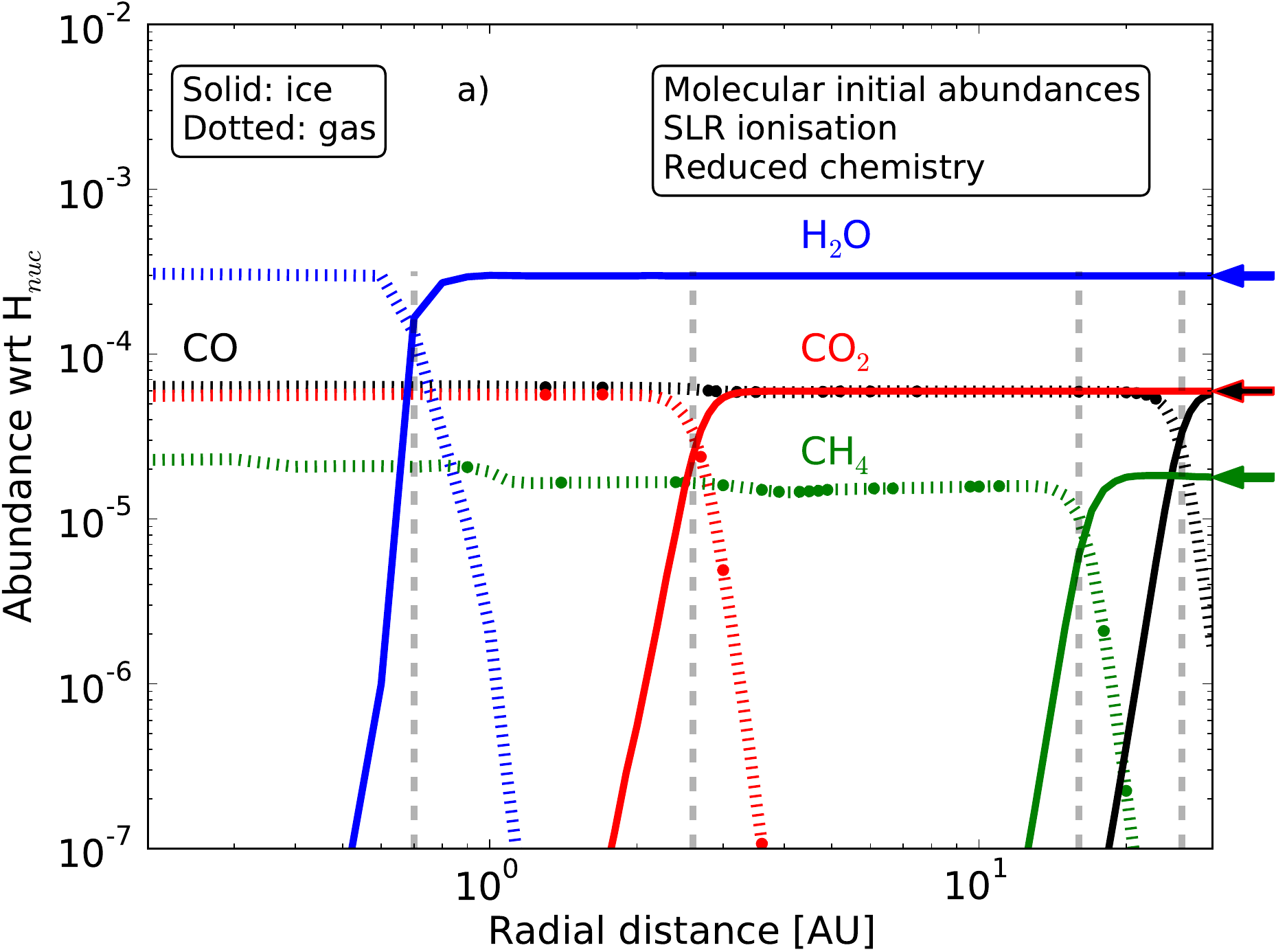}}
\subfigure{\includegraphics[width=0.5\textwidth]{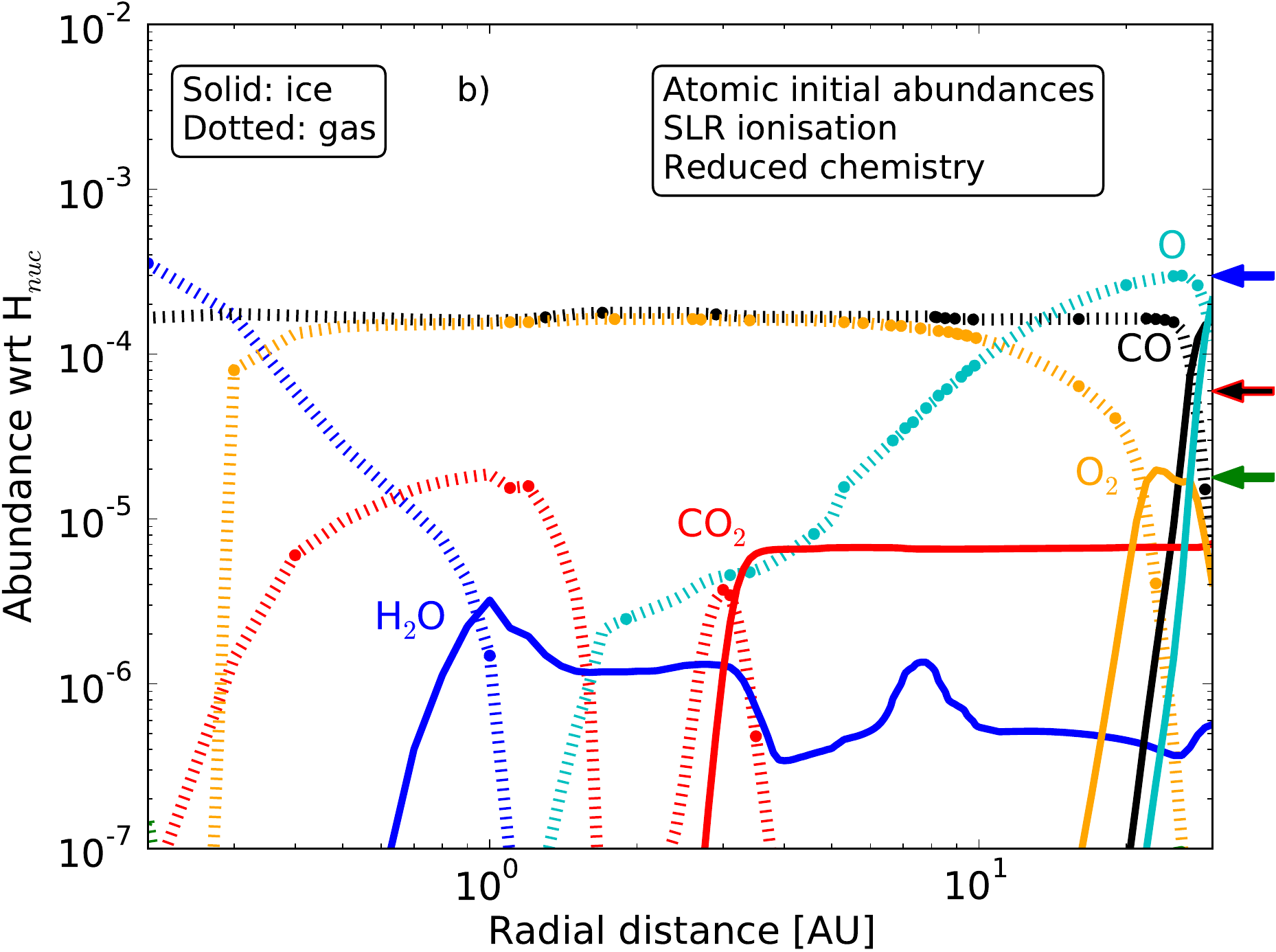}}\\
\subfigure{\includegraphics[width=0.5\textwidth]{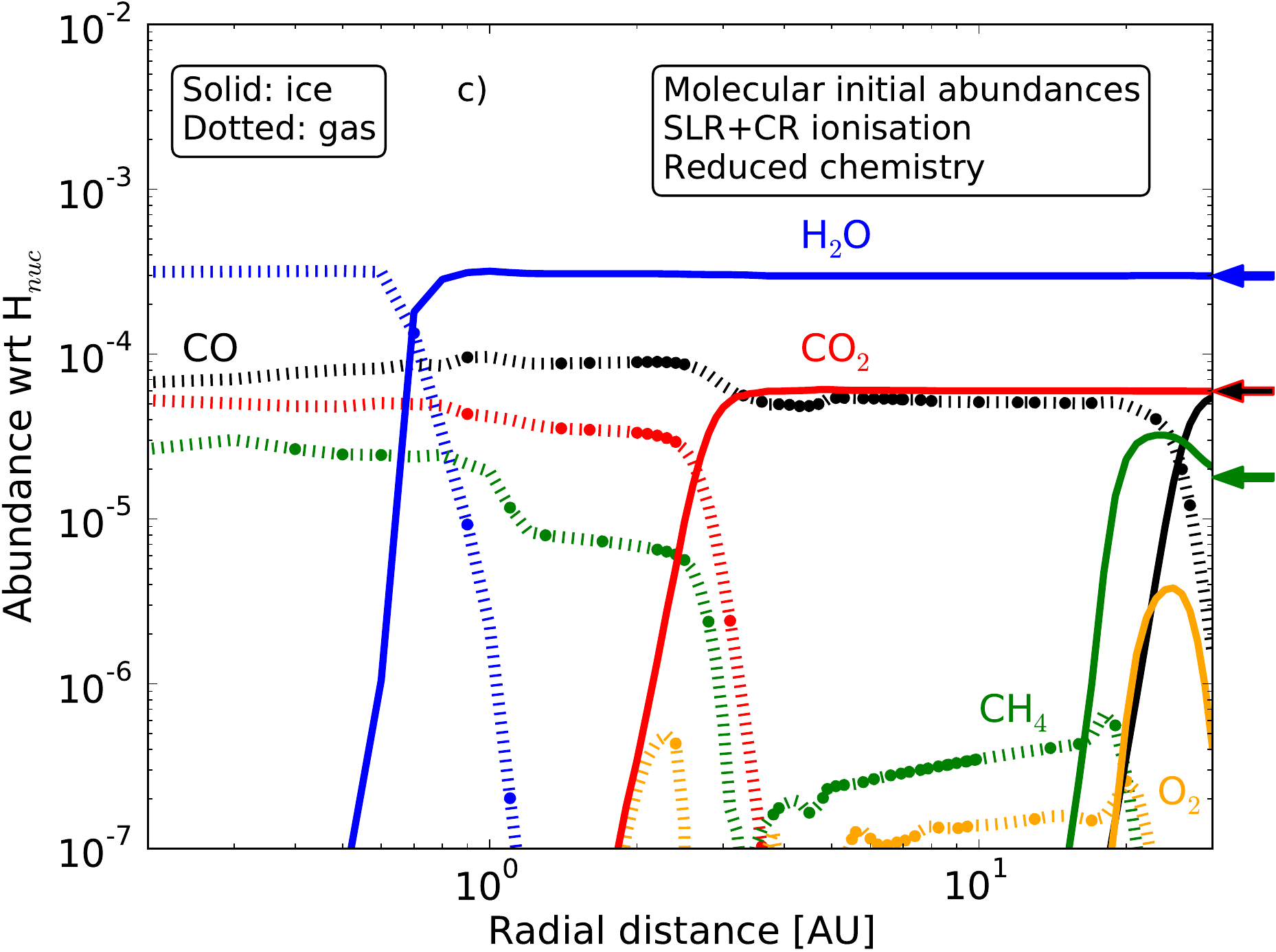}}
\subfigure{\includegraphics[width=0.5\textwidth]{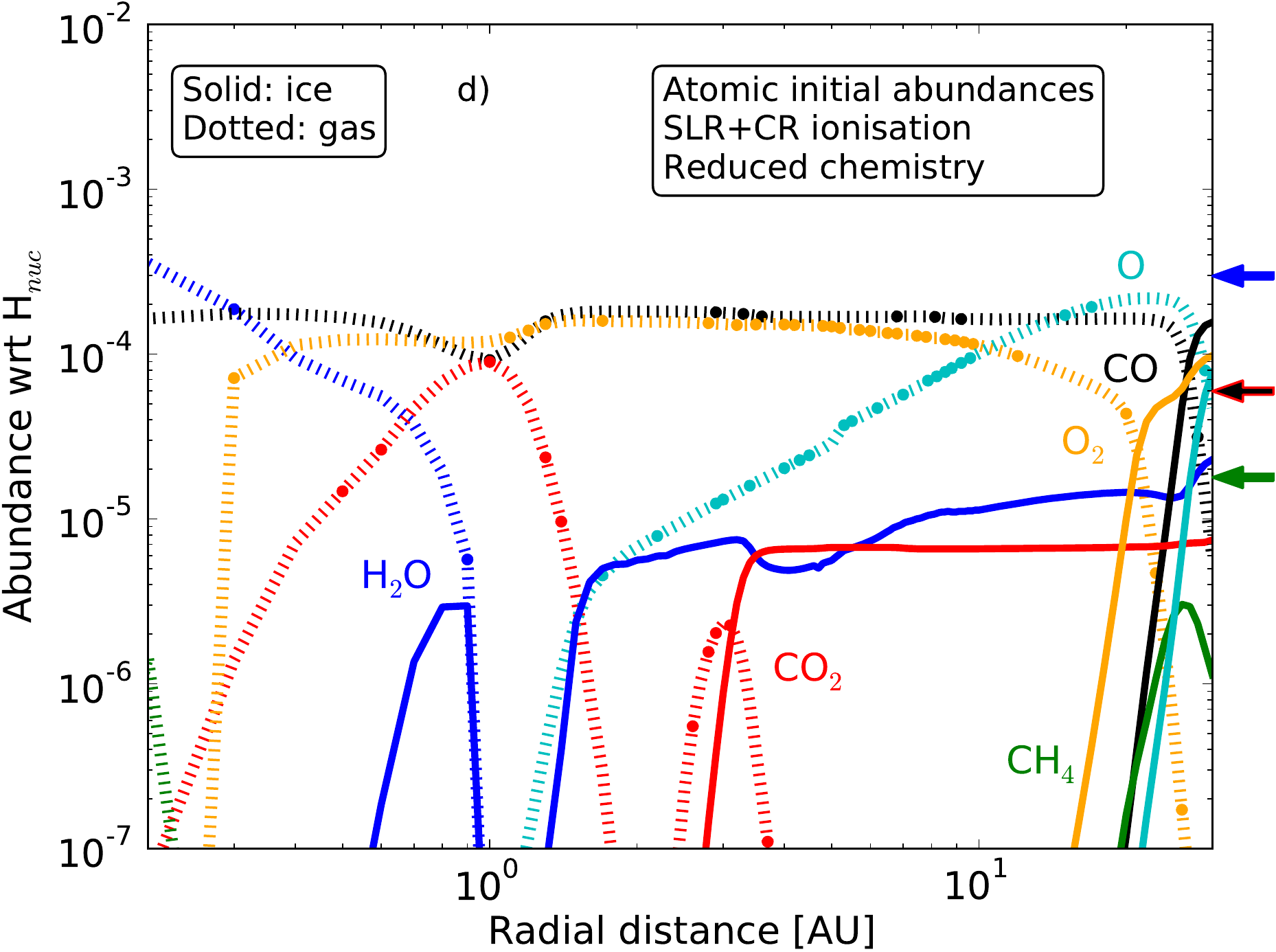}}
\caption{Final abundances with respect to total H nuclei density as function of radial distance $R$ from the 
star for key volatile species, when using the reduced chemical network (see Table~\ref{reac_types}). 
In all panels, the solid lines show the ice abundances and the dotted curves show the gas abundances. 
The top two panels show the results for the low ionisation case (SLRs only) 
and the bottom two panels show those for the high ionisation case (SLRs and CRs).
The left-hand panels show the results when assuming the reset scenario and the right-hand panels show those when assuming the reset scenario (see Table 1).
 (see Table~\ref{init_abun}). The arrows on the right-hand side of each plot indicate the initial abundances of \ce{H2O}, \ce{CO2}, CO, and \ce{CH4} gases in the inheritance scenario. CO and \ce{CO2} share the same arrow (red with black filling), because they have the same initial abundances. The grey, dashed, vertical lines in panel \textbf{a)} indicate the iceline positions of \ce{H2O}, \ce{CO2}, \ce{CH4} and CO, respectively, from the inner to the outer disk. The positions of these icelines are the same in the other panels.}
\label{simple_chem_plots}
\end{figure*}

Two different versions of the chemical network were utilised to 
obtain better insight into the importance of different processes: a reduced chemical network, 
and a full chemical network. 
The types of chemical reactions considered in the 
network versions are outlined in Table~\ref{reac_types}. 
The reduced chemical network comprises gas-phase chemical reactions, 
freezeout of gas-phase species onto grain surfaces to form ices, 
and desorption of ice species off grain surfaces back into the gas phase. 
The full chemical network also contains grain-surface chemistry in addition to
the reactions included in the reduced chemical network. 
For the reduced chemistry this means that a chemical species becomes
non-reactive as soon as it freezes out onto the surface of a grain, and that freeze-out and desorption of a species depend on the accretion and thermal desorption rates only. Photodesorption is excluded from the reduced chemical network to enable direct comparison to model results of \citet{marboeuf14}, whose only desorption mechanism is thermal desorption. For the
full chemistry, on the other hand, photodesorption is included, and chemical processing can continue
after freezeout. The motivation behind using a reduced and full chemical network, respectively, is to quantify the effects of gas-phase chemistry, and grain-surface chemistry, respectively.

The simulations are run with two different sets of initial abundances:
atomic species or molecular species (see Table \ref{init_abun} for an
overview of the initial species in each type of input). All abundances
in this paper are with respect to the total number of H nuclei. The
molecular abundances in Table \ref{init_abun} are values
representative of interstellar ices 
(see \citealt{oberg2011ices}, \citealt{boogert2015},
and Tables 1 and 2 in \citealt{marboeuf14}). For both sets of initial
abundances (atomic and molecular) the elemental ratios are consistent.

The choice of these initial abundances is motivated by the following
two scenarios about the history of the midplane material. The first
scenario is that the material going into protostellar systems is
inherited from the cloud out of which the protoplanetary disk
collapsed and formed. This scenario is denoted ``inheritance'', and it
implies that the material has the same composition as found in dark
clouds, especially their ices \citep[see][]{marboeuf14,charnley11}. 
The second scenario is the case
where the material coming from the dark cloud experiences heating
events from the protostar (i.e., accretion bursts or regular stellar
irradiation). These heating events are assumed to alter the chemistry
in disks significantly \citep[see, e.g.,][]{visser2015}. In the extreme
case, the chemistry is reset, meaning that the molecules are assumed to be dissociated into
atoms out to $R=$ 30 AU, which can then reform molecules and solids 
in a condensation sequence, as traditionally
assumed for the inner solar nebula \citep[e.g.][]{grossman1972}. Hence, the scenario considering atomic initial abundances is denoted
``reset''.  
Early chemical models of protoplanetary disks often assumed a set 
atomic initial abundances 
\citep[e.g.,][]{willacy1998,aikawa1999accre,semenov2004,vasyunin2008,walsh2010}; 
however, these early models also did not typically include a 
comprehensive grain-surface network and focussed solely on the gas-phase chemistry. Early models that did include grain-surface chemistry \citep[e.g.][]{willacy2007,walsh2010} were limited to simple atom-addition stemming from \citet{tielens1982} and \citet{hasegawa1993}. Here, a more comprehensive grain-surface network is used which includes radical-radical recombination, atom addition, and also ice processing \citep[][]{garrod2008}.
This work differs from earlier protoplanetary disk models 
in that we directly compare and quantify the effects of the 
inheritance versus reset scenarios for a single disk model 
using a comprehensive gas-grain chemical network.  

Important for investigating chemical evolution is also the size of the
grains in the disk midplane, which affects the rates of
grain-surface chemical reactions, as well as gas-grain interactions such as freezeout. 
Here, spherical grains are considered. Fixed sizes of
$r_{\rm grain} = 0.1 \mu {\rm m} = 10^{-5}$ cm and fixed grain number density
10$^{-12}$ with respect to H nuclei are assumed. 
Dust settling from the upper layers of a disk onto the disk 
midplane is assumed to happen on a timescale shorter than 1~Myr 
\citep{dullemond04,aikawa1999accre}, 
and this settling will increase the dust density relative to that for the gas. 
On the other hand, dust coagulation may decrease the dust surface area 
compared with that of standard ISM dust \citep{dullemond05}. 
The consequences of assuming fixed values are briefly discussed in 
Sect.~\ref{discussion}. 

All simulations are run for $t=1$~Myr. The lifetime of a disk before
gas dispersal is found to be in the range $<$ 1 to 10 Myr based on
observations \citep[see][]{williams11,fedele2011}. The time evolution
of the results is briefly discussed in Sects.~\ref{results} 
and \ref{discussion}.

\section{Results}
\label{results}

\begin{figure*}
\subfigure{\includegraphics[width=0.5\textwidth]{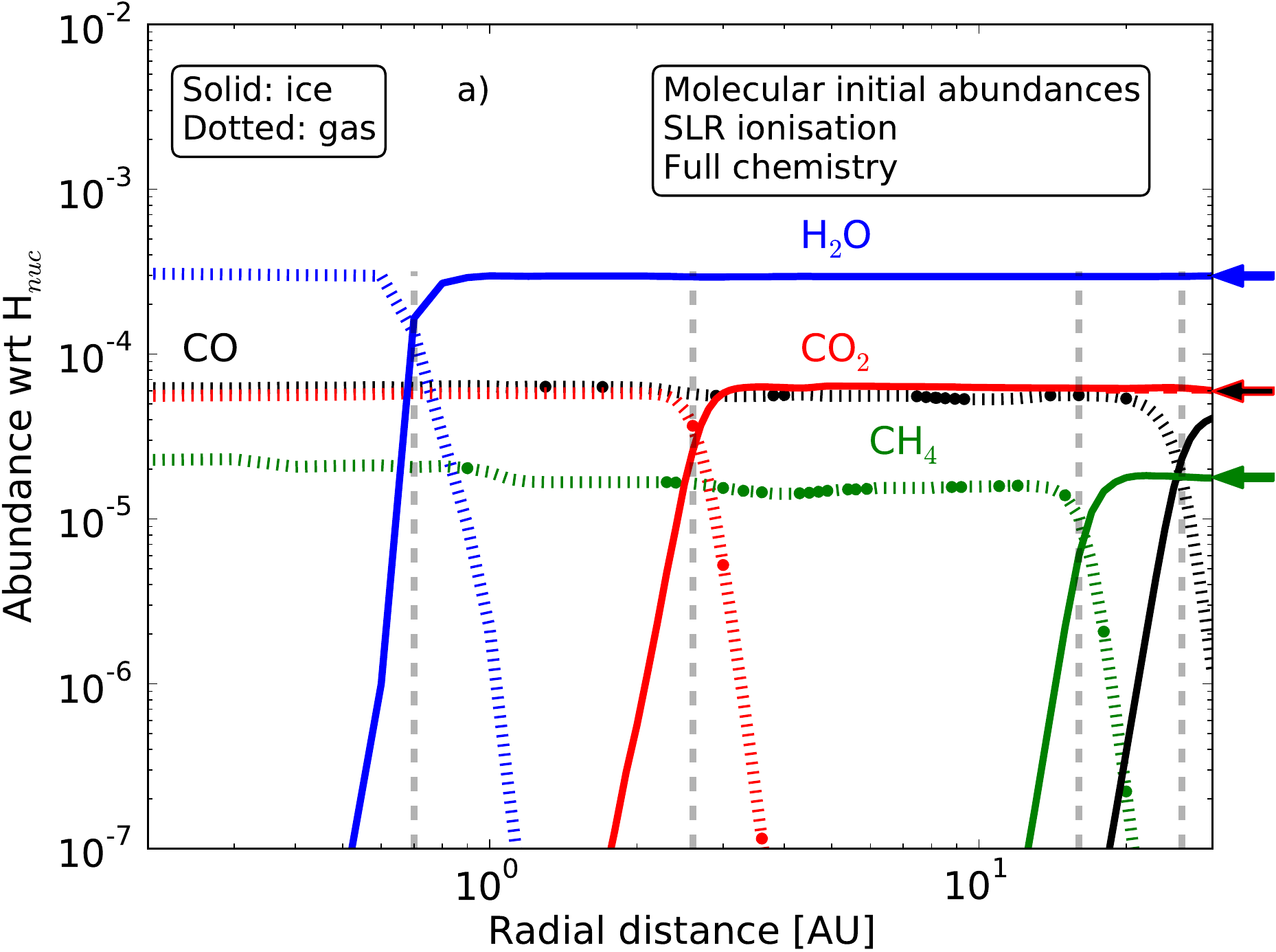}}
\subfigure{\includegraphics[width=0.5\textwidth]{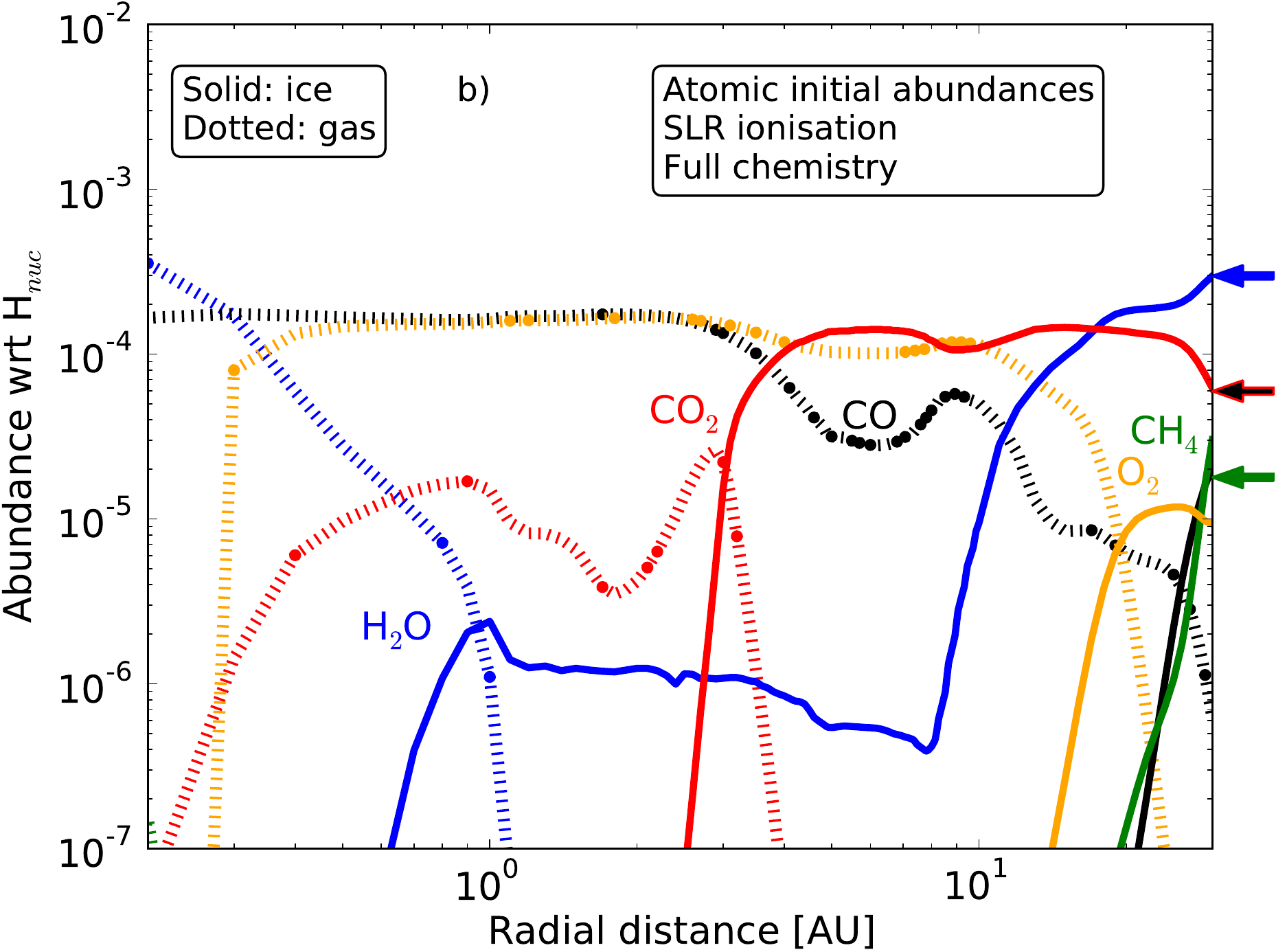}}\\
\subfigure{\includegraphics[width=0.5\textwidth]{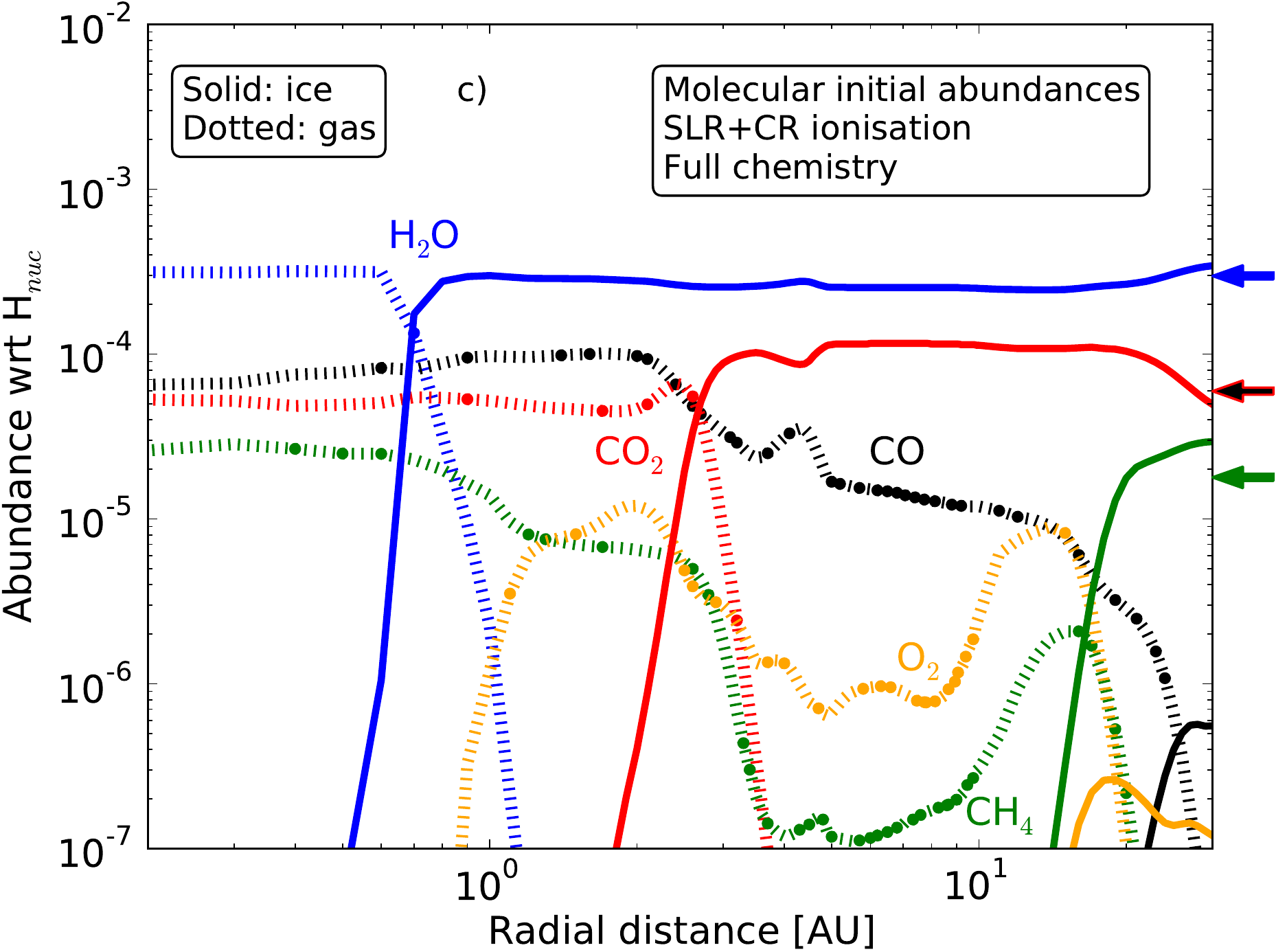}}
\subfigure{\includegraphics[width=0.5\textwidth]{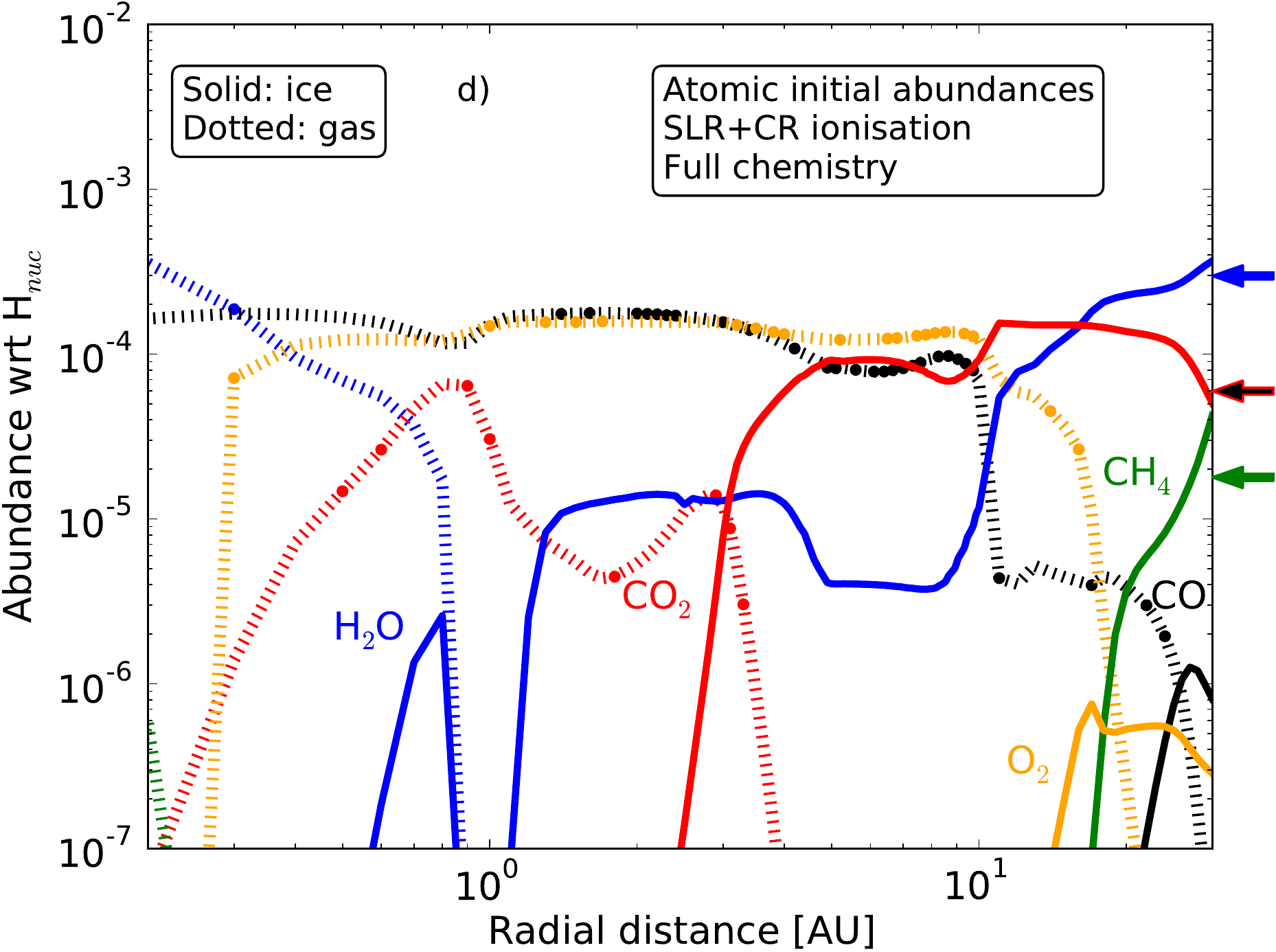}}
\caption{Final abundances with respect to total H nuclei density as function of radial distance from the 
star $R$ for key volatile species, when using the full chemical network (see Table~\ref{reac_types}). 
In all panels, the solid lines show the ice abundances and the dotted curves show the gas abundances. 
The top two panels show the results for the low ionisation case (SLRs only) 
and the bottom two panels show those for the high ionisation case (SLRs and CRs).
The left-hand panels show the results when assuming the reset scenario and the right-hand panels show those when assuming the reset scenario (see Table 1).
 (see Table~\ref{init_abun}). The arrows on the right-hand side of each plot indicate the initial abundances of \ce{H2O}, \ce{CO2}, CO, and \ce{CH4} gases in the inheritance scenario. CO and \ce{CO2} share the same arrow (red with black filling), because they have the same initial abundances. The grey, dashed, vertical lines in panel \textbf{a)} indicate the iceline positions of \ce{H2O}, \ce{CO2}, \ce{CH4} and CO, respectively, from the inner to the outer disk. The positions of these icelines are the same in the other panels.}
\label{full_chem_plots}
\end{figure*}

This section presents the results from the 
chemical evolution simulations for 
key volatiles which contribute significantly to the C/O ratio in the gas and ice: 
\ce{CO}, \ce{CO2}, \ce{H2O}, \ce{CH4}, \ce{O2}, O, HCN and NO (\ce{O2}, O, HCN and NO are presented only where relevant).
Chemical evolution results for the key nitrogen bearing species \ce{N2} and \ce{NH3} are also presented.
Fig.~\ref{simple_chem_plots} presents the results for the reduced chemistry, 
and Fig.~\ref{full_chem_plots} those for the full chemistry.
All figures show midplane abundances of chemical species with respect to 
\Hn~as a function of radial distance (in AU) from the central star. 
The arrows to the right of each panel indicate the initial abundance levels 
assumed for the inheritance scenario, see Table \ref{init_abun}. 
Colour coding of arrows matches that of the plots. 
CO and \ce{CO2} have the same assumed initial abundances.

In Fig.~\ref{simple_chem_plots}a, the icelines for each key volatile
considered here have positions at 0.7, 2.6, 16 and 26 AU, with temperatures of 177, 88, 28 and 21~K for \ce{H2O}, 
\ce{CO2}, \ce{CH4}, and CO, respectively (see icelines marked as dashed vertical lines in Fig. \ref{simple_chem_plots}a). 
The CO iceline is furthest out, because CO is more volatile than the other species, 
due to its low binding energy, see Table \ref{init_abun}. 
The positions of the icelines of these four volatiles are indicated with arrows in the top parts of each plot presented here.

\subsection{Reduced chemical network}
\label{simple_chem}

Figs.~\ref{simple_chem_plots}a and \ref{simple_chem_plots}c show the final abundances for 
the key volatiles when assuming molecular initial abundances, for 
the low ionisation and high ionisation case, respectively. 
This is the inheritance scenario.
Fig.~\ref{simple_chem_plots}a, assuming a low ionisation level, 
shows negligible changes in the ice abundances of each species 
(i.e., the initial molecular cloud abundances are preserved), 
and only minor changes to the gas-phase abundances within each iceline 
($<$10\% for CO and $<$30\% for \ce{CH4}). 

For the higher ionisation rate (see Fig.~\ref{simple_chem_plots}c), the
picture looks slightly different.  Larger changes in gas-phase
abundances are seen here for all the species inside their respective
icelines: up to 43\% for \ce{CO2}, up to 62\% for CO, and several
orders of magnitude for \ce{CH4}.  However, the \ce{H2O}, \ce{CO2},
and \ce{CO} ice abundances outside their respective icelines are
preserved with their initial assumed abundances.  This is due to the
almost instantaneous freezeout onto grains of these species under the
cold and dense physical conditions found in the outer disk, in
conjunction with the assumed chemical non-reactivity of these species
upon freezeout in the reduced form of the chemical network.

The largest difference in the gas-phase abundances in 
Fig.~\ref{simple_chem_plots}c when compared with 
those in Fig.~\ref{simple_chem_plots}a is the destruction of 
\ce{CH4} gas between $1 < R < 15$~AU, 
as well as the production of \ce{CH4} ice, reaching a peak abundance 
higher than the initial abundance level between $20$ and 25~AU, and also 
the production of \ce{O2} ice from $20$ to 30~AU. 
The differences in the outer disk can be ascribed to the increasing ionisation level here, as seen in Fig.~\ref{cr_level}.  
The higher ionisation rate creates a destruction pathway for abundant gas-phase species, such as \ce{CO}, 
which releases a small proportion of free C and O into the gas-phase for incorporation into 
other C- and O-bearing molecules, such as \ce{CH4} 
\citep[via the initiating ionisation-dependent reaction between CO and \ce{He+}, see][]{aikawa99} 
and \ce{O2}, via gas-phase reactions \citep[also discussed in][]{walsh15}.  
Beyond their respective icelines, \ce{CH4} and \ce{O2} 
freeze out onto grain surfaces almost instantaneously whereby they 
become depleted from the gas and are thus protected from further chemical modification.  
The gas-phase abundances of both species are not preserved within their 
respective icelines (16~AU for \ce{CH4} and 21~AU for \ce{O2})
because these species are also destroyed in the gas-phase by 
CR-induced photons (see reaction sequence \ref{cr_reactions}).   

Figs.~\ref{simple_chem_plots}b and \ref{simple_chem_plots}d show the results  
for the low and high ionisation cases (top and bottom panels, respectively) 
using atomic initial abundances. 
In this reset scenario, it is assumed that the gas has undergone 
an extreme heating event which has fully erased the prestellar composition 
(see Sect.~\ref{chem_model}).
For both ionisation levels, 
\ce{H2O} and CO are the dominant gas-phase species in the very innermost region of the 
disk midplane, $R<0.3$~AU.
For $R \geq 0.3$~AU, gas-phase CO and \ce{O2} are efficiently produced and 
reach similar abundance levels ($\sim 1.6\times 10^{-4}$ for \ce{O2} 
and $\sim 1.8\times 10^{-4}$ for CO)
out to their respective icelines at 21~AU for \ce{O2} and 26~AU for CO. 
Gas-phase \ce{O2} is much more abundant (by at least 2 orders of magnitude) 
in this reset scenario than in the inheritance scenario. In the reset scenario \ce{O2} is formed is the gas-phase via \ce{O + OH -> O2 + H} \citep[see][]{walsh15}, and remains in the gas-phase because it is very volatile (binding energy of $E_{b}=1000$ K), and only freezes out at 24 K, see Table \ref{init_abun}.
Outside the CO iceline, the CO ice abundance reaches the same level 
as that for the gas inside the iceline. 
This CO abundance is about half the \ce{H2O} initial abundance  
in the inheritance scenario. 
The remainder of the available oxygen (63\%) in the outer disk remains in atomic form.   
These are oxygen atoms that have not had sufficient time to form molecules before freezeout. 
30~AU is just around the atom oxygen iceline (which is at 29 AU), 
so both O gas and ice account for the total O abundance of $3.3\times10^{-4}$, 
making atomic oxygen the most abundant O carrier here, 
almost twice as abundant as CO ice. 
This result is particular to the reduced chemistry case because all species 
are rendered chemically inert upon freezeout; in reality, 
atomic oxygen on and within ice mantles is highly reactive 
\citep[see, e.g.,][]{linnartz2015}.  

Between $R =0.2$ and 1.6~AU, gas-phase \ce{CO2} is produced for both ionisation levels. 
The abundance reaches 9.0$\times 10^{-5}$ for the high ionisation case at $R=1$~AU 
(Fig. \ref{simple_chem_plots}d), which is about 5 times higher than the abundance peak for 
low ionisation at $R=1$ AU (see Fig. \ref{simple_chem_plots}b). 
For both low and high ionisation rates, \ce{CO2} ice reaches an abundance of only about
$10$\% of the value assumed in the inheritance scenario outside 
the \ce{CO2} iceline.
In addition, \ce{H2O} ice is not efficiently produced in the outer disk in this reset 
scenario. The reset scenario \ce{H2O} abundance levels resemble abundance levels of \ce{H2O} naturally produced in the gas phase of $10^{-7}-10^{-6}$ \citep[see][]{hollenbach2009}.
The abundance of \ce{H2O} ice is about an order of magnitude larger in the 
high ionisation case, showing that ion-molecule reactions in the gas-phase are 
contributing to the formation of water in the absence of grain-surface chemistry.  
However, the peak \ce{H2O} ice abundance of 2.3$\times 10^{-5}$ 
(reached at 30~AU in Fig.~\ref{simple_chem_plots}d) remains more than an order of 
magnitude lower than the initial \ce{H2O} abundance assumed in the inheritance scenario.
The high ionisation level in Fig.~\ref{simple_chem_plots}d also aids the formation of \ce{CH4} 
ice beyond its iceline, reaching a peak abundances of 3.0$\times 10^{-6}$.  
However, as also found for \ce{H2O} and \ce{CO2} ice, the maximum abundance 
reached for \ce{CH4} is only 17\% of the assumed initial abundance for the 
inheritance scenario.

\subsection{Full chemical network}
\label{full_chem}

Fig.~\ref{full_chem_plots}a shows the chemical evolution results when assuming the 
inheritance scenario and a low ionisation rate (SLRs only) for the full chemical network. 
The abundance behaviour in this figure is practically indistinguishable 
from that for the case using the reduced chemical network (see Fig.~\ref{simple_chem_plots}a), 
that is, the initial assumed abundances are preserved in both the gas
and the ice.  
For the higher ionisation rate case (SLRs and CRs, Fig.~\ref{full_chem_plots}c) 
the picture is different.  
Chemical processing by cosmic-ray-induced reactions in both gas and ice occurs,
and is most noticeable within the icelines of each species.  
A large reduction in the gas-phase CO abundance beyond
2~AU is seen, in contrast with the reduced chemistry results. 
This decrease in CO gas (and ice beyond the iceline) 
coincides with an overall enhancement in \ce{CO2} ice in the outer disk. 
CO molecules accreting onto the grain surfaces can react with OH
radicals produced in the ice via photodissociation of \ce{H2O} ice by
CR-induced photons in the full chemistry.  
This produces \ce{CO2} ice in-situ on the grain surfaces via the reaction
\beq \ce{CO_{surf}} + \ce{OH_{surf}} \longrightarrow \ce{CO2_{,surf}}
+ \ce{H_{surf}},
\label{co2_prod}
\eeq 
which is responsible for the rise in the \ce{CO2} ice abundance
between 3 and 15~AU (doubling the initially assumed \ce{CO2}
abundance).  
Within this radial region the dust temperature is between
27~K and 78~K, and because H atoms are very volatile ($E_{\rm b} = 600$~K) 
they can rapidly thermally desorb from the grain surface, impeding
further grain-surface reactions involving atomic H.  
In the outermost disk between 20 and 30~AU, the temperature drops to 
$T\approx 19.5$~K, enabling the more efficient retention of H atoms 
arriving from the gas or produced in-situ within the grain mantles, 
and thereby increasing the relative rate of \ce{H2O} ice production via the reaction, 
\beq 
\ce{OH_{surf} + H_{surf} <=>[][\gamma_\mathrm{CR}] H2O_{surf}}.
\label{h2o_prod}
\eeq 
This reaction is more efficient than Reaction~(\ref{co2_prod})
under these colder conditions. Although balanced by destruction due to
photodissociation, the abundance of \ce{H2O} ice is increased at the
expense of \ce{CO2} ice beyond $20$~AU.

\begin{figure}
\includegraphics[width=0.5\textwidth]{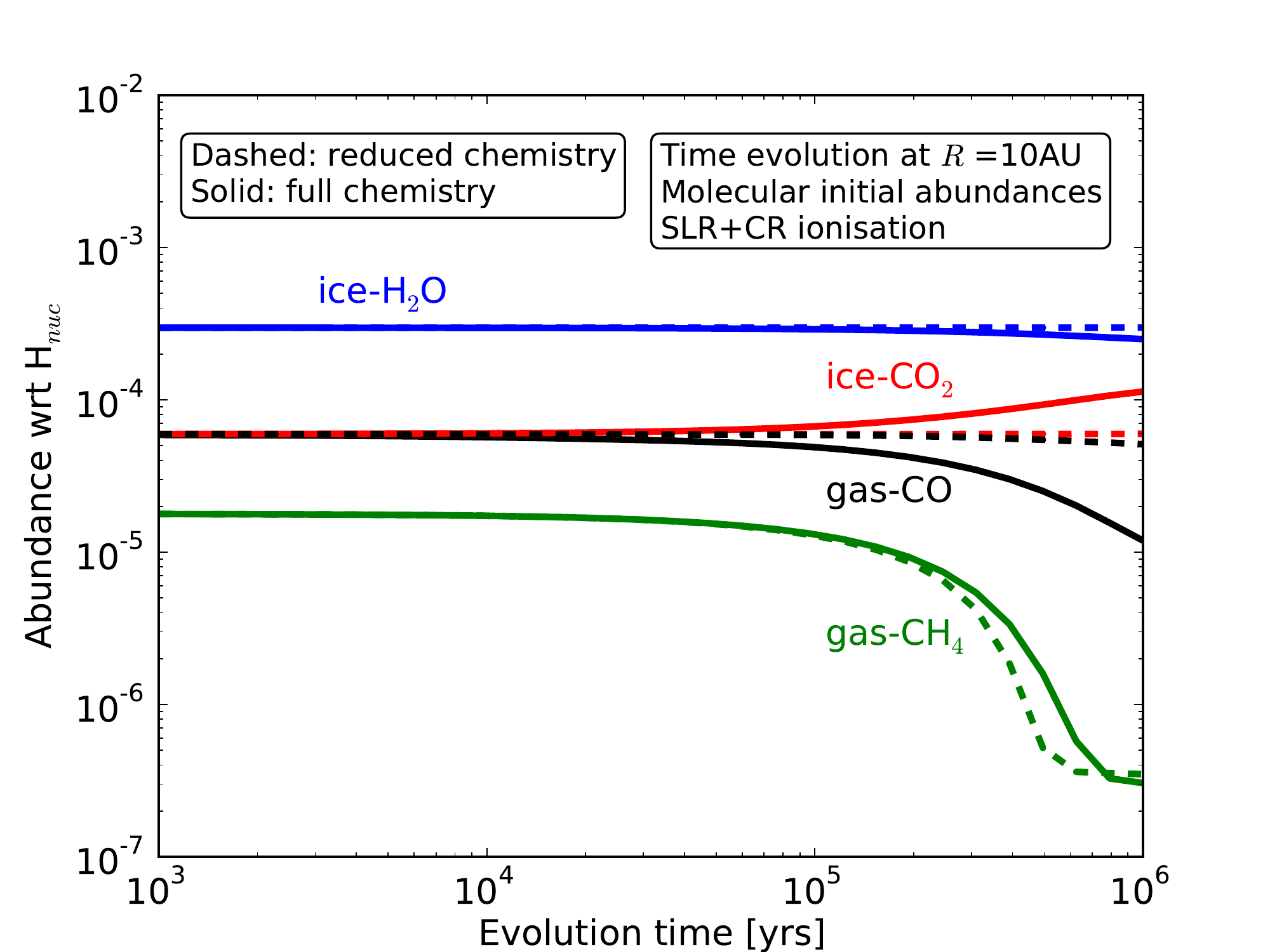}
\caption{Abundance evolution with time for CO gas, \ce{CO2} ice, 
\ce{H2O} ice and \ce{CH4} gas. Dashed curves are for the reduced 
chemistry, solid curves are for the full chemistry. Additional time-dependent plots are supplied in Appendix \ref{timescale-app}}
\label{meth_evol}
\end{figure}

\ce{CH4} gas is destroyed between 1 and 16~AU in the full chemistry, 
similar to the case for the reduced chemistry (see Fig.~\ref{simple_chem_plots}c).
The reactions responsible for this destruction of
\ce{CH4} at 10~AU, are
\begin{align}
\ce{CH4} & \ce{->[\gamma_\mathrm{CR}] CH2} \ce{->[\ce{H}] CH3} \ce{->[\ce{O}] H2CO},\\
\ce{CH2} & \ce{->[\ce{NO}] HCNO} \ce{->[\ce{O}] CO} \ce{->[\ce{OH}] CO2},
\end{align}
where $\gamma_\mathrm{CR}$ indicates photodissociation by cosmic-ray
induced photons.
The carbon is thus converted from \ce{CH4} into \ce{CO2} and 
\ce{H2CO}, but the conversions require CRs (i.e. high ionisation). 
However, \ce{H2CO} does not reach an abundance of 
more than 10$^{-12}$, so \ce{CO2} is the main reservoir for the 
carbon converted from the \ce{CH4} gas, see Fig. \ref{meth_evol}. Some carbon from \ce{CH4} is also converted into unsaturated hydrocarbons in the gas phase \citep[as also seen in][]{aikawa1999accre}, eventually forming gas-phase \ce{C2H4}. This is the reservoir for the converted carbon in the reduced chemistry, because \ce{CO2} ice requires gran-surface chemistry to form, and hence \ce{CO2} ice does not increase in abundance in the reduced chemistry case. For the full chemistry, however, this \ce{C2H4} subsequently freezes out onto grain surfaces, undergoes hydrogenation, and forms \ce{C2H6} which reaches a final abundance of 8.6$\times 10^{-6}$, approximately an order of magnitude lower than that of \ce{CO2} (1.2$\times 10^{-4}$).

The reaction \ce{CO + He+} reported by \citet{aikawa1999accre}, turning carbon from CO into \ce{CH4} and other hydrocarbons, is found here to be a relatively minor channel as compared to the reactions between \ce{CH4} and \ce{He+}, and \ce{H3+}, respectively, turning \ce{CH4} into CO. The result of this is an efficient conversion of carbon from \ce{CH4} into CO. Outside the \ce{CO2} iceline, this conversion is enhanced. This enhancement is due to the accretion of CO onto grain surfaces, and subsequent rapid reaction of CO with OH to form \ce{CO2}, a reaction not included in the models of \citet{aikawa1999accre}.

The destruction of CO gas due to the conversion into \ce{CO2} has been suggested by e.g.  \citet{nomura2016} to explain the depletion of CO gas inside the CO iceline of the protoplanetary disk TW Hya. \citet{kama2016codepl} reported overall carbon depletion in the disk atmospheres of some protoplanetary disks using results from a single-dish survey of CO ($J=6-5$) and [CI]($^{3}P_{1}-^{3}P_{0}$) line emission with APEX. \citet{schwarz2016} reported CO depletion in TW Hya from CO isotopologue emission observations with ALMA. Detailed modelling confirmed that carbon is likely depleted in the disk around TW Hya by a factor of $\approx$ 100 (\citealt{kama2016model}, see also \citealp{du2015}). This is in good agreement with the CO gas abundance just inside the CO iceline in Fig. \ref{full_chem_plots}c, and the results presented here therefore provide a possible explanation for the presence of this inner-disk CO depletion, found inside of, but somewhat mimicking, the actual CO iceline in TW Hya.

At a temperature of 37.7~K (at $R=10$~AU), \ce{CO2} freezes out immediately after production. 
The abundance increase in \ce{CO2} ice follows the destruction 
of \ce{CH4} gas, as seen in Fig.~\ref{meth_evol} which presents the time 
evolution. 
It is interesting that this effect is seen under these specific 
physical conditions, with temperatures ranging from 40 to 150 K. 
This indicates that the \ce{CH4} destruction happens at radii between 
the \ce{H2O} and \ce{CH4} icelines, where 
\ce{H2O} is not chemically active in gas-phase reactions. More generally, for the high ionisation level, the chemical processing becomes significant after a few times $10^{5}$~yrs. That is shown in Appendix \ref{timescale-app}, where jumps in abundance levels are presented for our four key volatiles between 100 kyr and 500 kyr.

When considering initial atomic abundances (Figs.~\ref{full_chem_plots}b
and \ref{full_chem_plots}d) a similar radial behaviour (although not
identical) is seen when comparing the results for reduced and full
chemistry.  Inside the icelines, mainly gas-phase \ce{H2O}, \ce{O2}
and CO are produced, reaching peak abundances of $3.6 \times 10^{-4}$,
$1.6 \times 10^{-4}$, and $1.8 \times 10^{-4}$, respectively.  
Not as much gas-phase \ce{CO2} is produced within the iceline. 
Comparing the results for low and high ionisation, the peak abundances of gaseous \ce{CO2} 
are $2.5 \times 10^{-5}$ and $6.6 \times 10^{-5}$, respectively, 
and a negligible amount of gas-phase \ce{CH4} is formed.  For the high ionisation
rate, Fig.~\ref{full_chem_plots}d, a larger amount of \ce{H2O} ice (an
order of magnitude at 2~AU) is produced between 1 and 10~AU, than with
a low ionisation rate (Fig.~\ref{full_chem_plots}b), as is also seen in
the model using the reduced chemical network. For \ce{H2O} ice, a dip in the abundance is seen around 7 AU in Fig.~\ref{full_chem_plots}d. This is an effect of the competing productions of \ce{H2O} ice and \ce{SO2} ice. At 4 AU, \ce{SO2} ice is produced more slowly than \ce{H2O} ice, and reaches a final abundance of 2.7$\times 10^{-6}$. At 7 AU, however, the lower temperature favours a more efficient and fast production of \ce{SO2} ice, which is able to lock up atomic O, thereby impeding the simultaneous production of \ce{H2O} ice.

Can these models produce abundant O$_2$ in gas or ice?  Significant
gaseous \ce{O2} is formed in the inner disk starting from atomic
abundances, as discussed above. Using the low ionisation rate (see
Fig. \ref{full_chem_plots}b), \ce{O2} ice reaches an appreciable
abundance of 1--10\% that of \ce{H2O} ice between 15 and 30~AU,
similar to that seen in the results using the reduced chemical
network.  However, little \ce{O2} ice is found when using the higher
ionisation rate, indicating that \ce{O2} ice is susceptible to
chemical processing by CR-induced photons.  In
Fig.~\ref{full_chem_plots}c using molecular initial abundances, \ce{O2}
gas is also produced from 1~AU out to 15~AU but in smaller amounts
than starting from atomic initial abundances. It reaches a peak
abundance of a few percent of that of \ce{H2O} ice at 2~AU.

\subsection{Main nitrogen reservoirs}
\label{nitrogen}

\begin{figure*}
\subfigure{\includegraphics[width=0.5\textwidth]{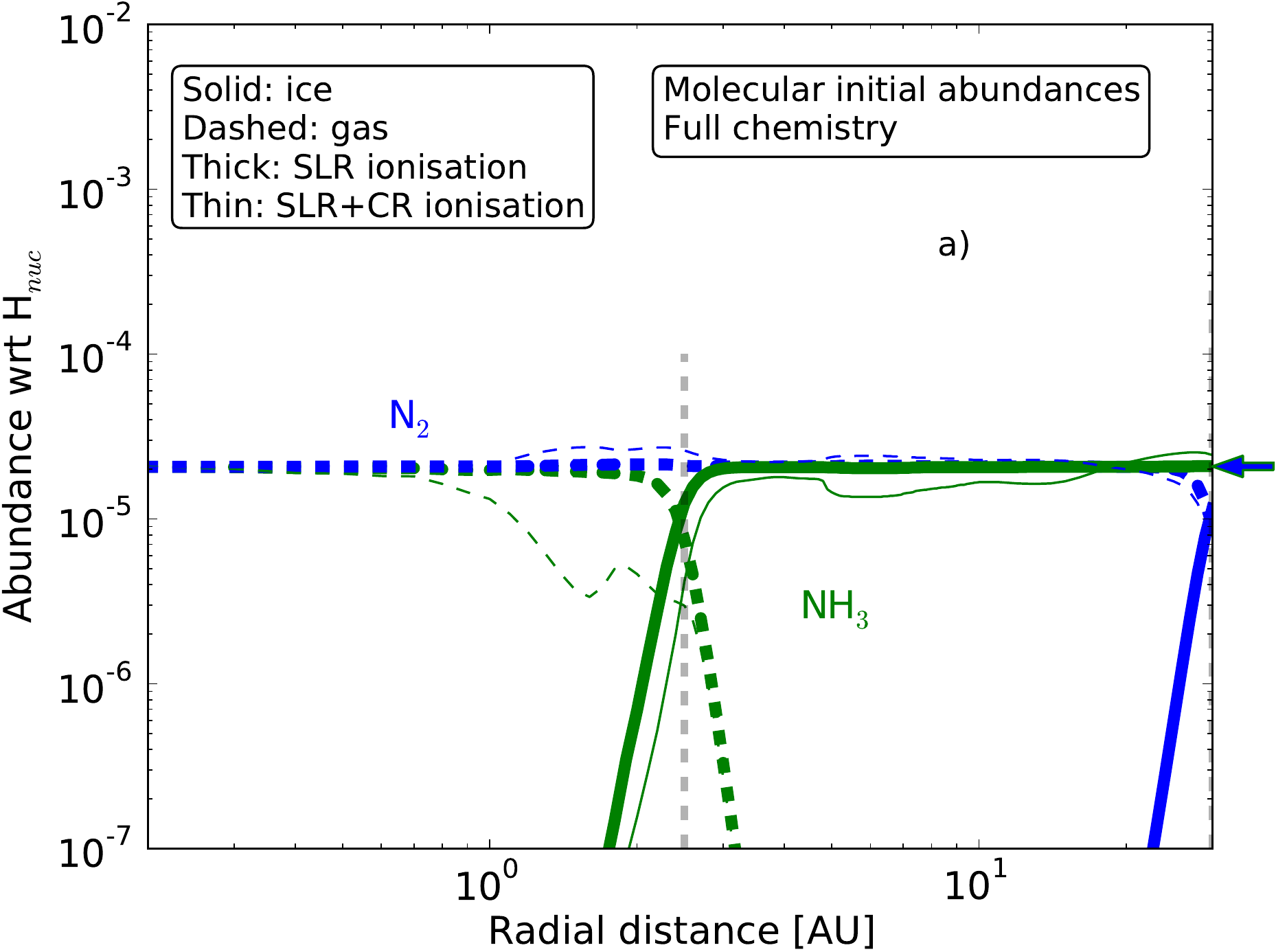}\label{nitro_mol}}
\subfigure{\includegraphics[width=0.5\textwidth]{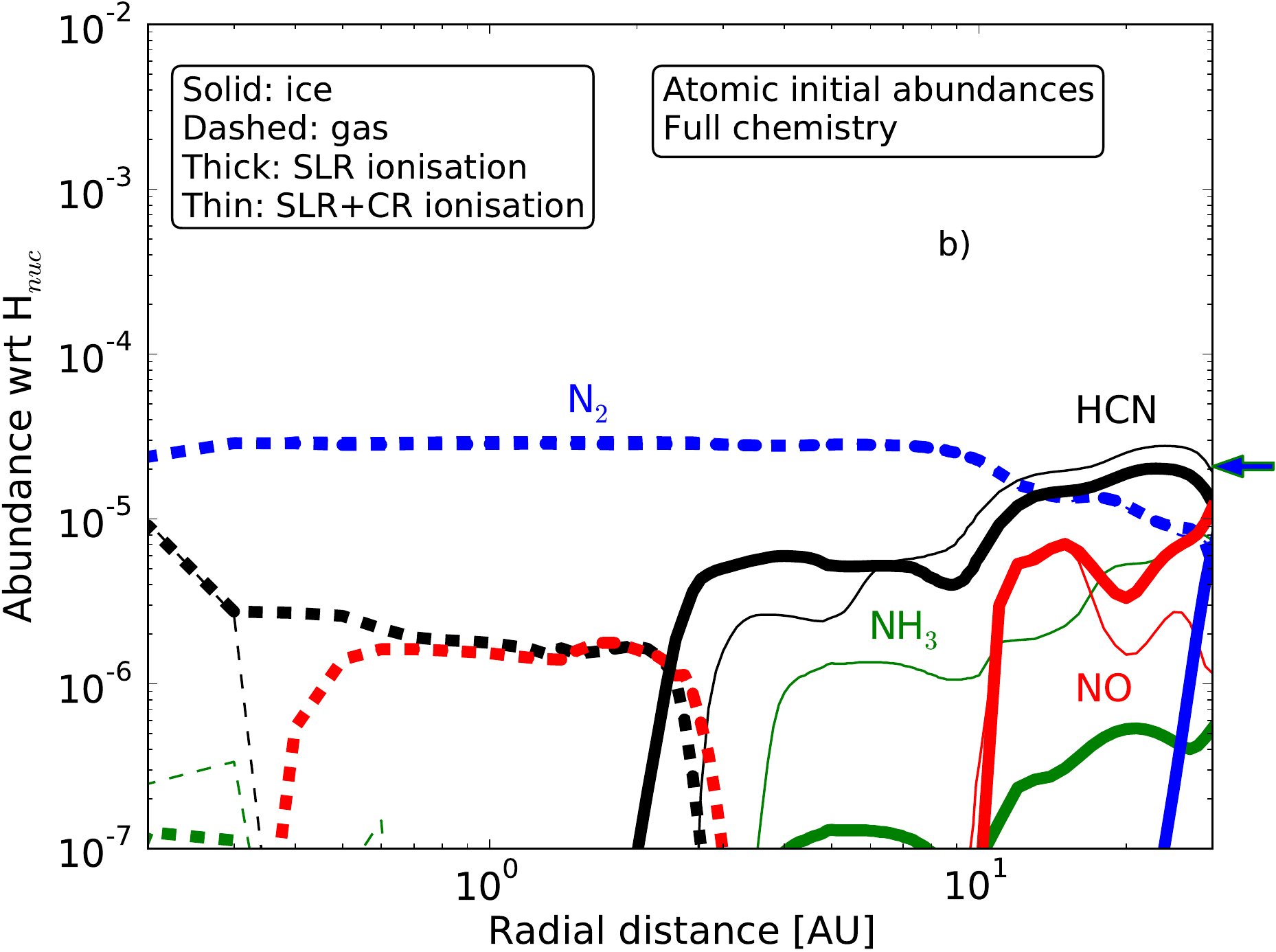}\label{nitro_atom}}
\caption{Final abundances for the major nitrogen species obtained with the full chemical network. 
Solid curves show ice abundances, dashed curves show gas abundances. 
Thin curves are for low ionisation level, and thick curves are for high ionisation level. 
The physical conditions and 
initial abundances in Fig.~\ref{nitros}a are identical to those 
considered in Fig.~\ref{full_chem_plots}a and c. 
Likewise, for Fig.~\ref{nitros}b 
the assumptions are identical to those in Fig.~\ref{full_chem_plots}b and d. The arrow on the right-hand side of each plot indicate the initial abundances of \ce{N2} and \ce{NH3}. They have the same initial abundances and hence share arrow. The grey, dashed, vertical line in panel \textbf{a)} indicates the iceline position of \ce{NH3}. The position of the iceline is the same in panel \textbf{b)}. The iceline of \ce{N2} is at 30 AU, and therefore not seen.}
\label{nitros}
\end{figure*}

Figs.~\ref{nitros} shows the final abundances for \ce{N2}, 
\ce{NH3}, HCN and NO (HCN and NO only where relevant) for the full chemistry. 
Results for both low and high ionisation levels, 
and both gas and ice species, are plotted in each figure. 
Fig.~\ref{nitros}a are the results when assuming molecular initial abundances 
(i.e. the inheritance scenario), whereas 
Fig.~\ref{nitros}b are those assuming atomic initial abundances 
(i.e. the reset scenario). 
For the inheritance-scenario with low ionisation, 
the initial abundances are largely preserved, as for the C- and O-bearing species, and the icelines are of \ce{NH3} and \ce{N2} are nicely outlined at 2.5 and 30 AU, at temperatures of 90 and 20 K, respectively. These icelines are marked with dashed vertical lines in Figs. \ref{nitros}a. 
However, for high ionisation there is a destruction of 
\ce{NH3} and production of \ce{N2} at $\approx 1.5$~AU. 
The reaction pathways responsible for these features are 
as follows:

\begin{align}
\ce{NH3} &\ce{ ->[X\ce{H+}] NH4+ ->[e-] NH2 ->[\ce{NO}] N2}  \label{nitro_ion1}\\
\ce{NH3} &\ce{ ->[\ce{OH}] NH2 ->[\ce{NO}] N2}  \label{nitro_oh}\\
\ce{NH3} &\ce{ ->[\gamma_{CR}] NH2 ->[\ce{NO}] N2}  \label{nitro_ion3} 
\end{align}

\ce{NH3} is converted into \ce{N2} both through ion-molecule reactions 
(reaction sequence \ref{nitro_ion1}), 
as well as through CR-induced photoreactions 
(reaction sequences \ref{nitro_oh} and \ref{nitro_ion3}). The timescale of this conversion is a few times $10^{5}$ yrs, as was found for the cases of \ce{H2O}, CO, \ce{CO2} and \ce{CH4} and discussed earlier in Section \ref{full_chem}.

When considering atomic initial abundances in Fig.~\ref{nitros}b, 
atomic N is seen to form \ce{N2} gas quickly 
\citep[as also shown and discussed in][]{schwarz2014}, which is the dominant bearer inside 10~AU. 
This holds regardless of the assumed ionisation level.  
In the outer disk, HCN ice is the main reservoir, being 5-10 times more
abundant than \ce{NH3} ice outside 10~AU. 
For low ionisation around 1~AU, NO and HCN are the second and third 
most abundant N-bearing species, although more than an order of magnitude 
less abundant than \ce{N2}. 
In the outer disk, \ce{NH3}, \ce{N2}, and NO all reach ice
abundances factors of a few to ten times less than the HCN ice abundance.
HCN is produced on very short timescales ($< 1$~yr) 
in the gas-phase via the following reaction sequence, 
\beq
\ce{C ->[\ce{H2}] CH2 ->[\ce{H2}] CH3 ->[\ce{O}] HCO ->[\ce{N}] HCN},
\eeq 
with subsequent freeze-out of HCN onto grains. 
NO is produced mainly via the well-known gas-phase reaction between N and OH, 
whereas \ce{NH3} ice is formed in situ on the grain surfaces 
via hydrogenation of atomic N \citep[see also][]{walsh15}. 
Even with grain-surface chemistry included, when beginning with atomic 
initial abundances, \ce{NH3} ice formation is less efficient than that 
for those species more reliant on gas-phase formation followed by freezeout.  

\section{Discussion}
\label{discussion}

\subsection{Compositional diversity at different radii: inheritance vs reset}
\label{comp_div}


The third column of Table \ref{result_composition} shows the final 
abundances for the full chemical model with low ionisation level and molecular 
initial abundances at three different radii throughout the disk midplane 
(see Fig. \ref{full_chem_plots}a).  
This is the model for which the results show best preservation of the 
initial abundances (listed in Table~\ref{init_abun} for the inheritance case).
The remaining columns show the 
fractional abundances of each species at each radial distance for all 
other models using full chemistry relative to this reference model. The fractional abundances have been calculated using the formula given in the footnote of Table \ref{result_composition}.
Fractional abundance values smaller than 1 means abundances lowered relative 
to the reference model whereas values larger than 1 means increased abundances.  
Values of 0 indicate abundances 
more than two orders of magnitude lower than for the reference model, so as to avoid large numbers in the table.  
A value of 1 means no change from the reference model.  
Three different radii are considered which span the radius of 
the disk and probe three distinct regions: 
(i) \ce{H2O}-ice rich only (1~AU), 
(ii) \ce{H2O}-ice rich and \ce{CO2}-ice rich (10~AU), and  
(iii) volatile-gas poor (30 AU, i.e., gas fully depleted of volatiles via freezeout).  

\begin{table*}
\renewcommand{\arraystretch}{1.2}
\caption{Fractional deviations in key volatiles between different simulations 
using full chemistry 
at $R=1$, 10, and 30~AU with respect to the reference model 
(the abundances for which are given in the column labelled ``Mol. low ion.''). }
\centering                          
\begin{tabular}{l l c c c c}        
\hline\hline
Radial distance & Species & Abundance& & Fractional deviation & \\ 
\hline
& &(Reference model) & & & \\
& & Mol. low ion.& Mol. high ion. & Atom. low ion. & Atom. high ion.\\
\hline
 $R$ = 1~AU   &   \emph{Gas} &&&&\\      
&   \ce{H2O}  & 2.2$\times 10^{-6}$    & 1.03    &0.5   &0 \\
&   \ce{CO}   & 6.4$\times 10^{-5}$    & 1.52   &2.5  &2.4 \\
&   \ce{CO2}  & 5.8$\times 10^{-5}$    & 0.9   &0.26   &0.53  \\
&   \ce{CH4}  & 1.9$\times 10^{-5}$    &0.74    &0  &0 \\
&   \emph{Ice} &&&&\\      
&   \ce{H2O}  & 3.0$\times 10^{-4}$    & 1  &0  &0 \\               
\hline            
 $R$ = 10~AU  &   \emph{Gas} &&&&\\      
&   \ce{CO}    & 5.3$\times 10^{-5}$     &0.23  &0.95   &1.31 \\
&   \ce{CH4}   & 1.6$\times 10^{-5}$     &0.02 &0 &0 \\ 
&   \emph{Ice} &&&&\\      
&   \ce{H2O}   & 3.0$\times 10^{-4}$     & 0.83  &0.04  &0.04 \\
&   \ce{CO2}   & 6.3$\times 10^{-5}$     &1.79   &1.73  &1.46 \\
\hline 
 $R$ = 30~AU  &   \emph{Gas} &&&&\\      
&   \ce{CO}   & 1.2$\times 10^{-6}$      & 0.01   &0.53  &0.02\\
& \emph{Ice}  &&&&\\      
&   \ce{H2O}  & 3.0$\times 10^{-4}$      & 1.14  & 1   &1.22\\
&   \ce{CO}   & 4.1$\times 10^{-5}$      &0.01   &0.5  &0.02\\
&   \ce{CO2}  & 6.0$\times 10^{-5}$      &0.83   &1.04   &0.84\\
&   \ce{CH4}  & 1.8$\times 10^{-5}$      &1.67   &1.71  &2.36\\ 
\hline
\label{result_composition}
\end{tabular}\tablefoot{  
Formula for calculating fractional deviations: deviation=abundance(comparison model)/abundance(reference model).\\
The reference model, ``Mol.~low ion.'', is that for which the initial abundances are mainly preserved 
throughout the midplane.  
In all subsequent column labels, ``Mol'' and ``Atom'' refer to molecular and 
atomic initial abundances, and ``low'' and ``high'' refer to a 
low assumed ionisation rate (SLRs only) and a high ionisation rate 
(SLRs and CRs), respectively.}
\end{table*}


At $R=1$~AU, CO gas is enhanced in all three models 
by factors of ~1.5 to 2.5, with a greater enhancement  
seen in the atomic case.  
This is generally at the expense of all other 
considered species: \ce{H2O} ice and gas, \ce{CH4} gas, and 
\ce{CO2} gas. 
Both \ce{CH4} gas and \ce{H2O} ice show extreme depletion for 
the atomic cases, for the reasons discussed in Sect.~\ref{results}.  

Moving to 10~AU, i.e., beyond the \ce{CO2} iceline, a different 
behaviour is seen. 
\ce{CO2} ice is enhanced by factors of ~1.5 to 1.8 in the 
other three models, this time generally at the expense of \ce{CO} 
and \ce{CH4} gas and \ce{H2O} ice.  
The extreme depletion of water ice in the atomic cases 
also extends into this region.  
However, an enhancement of a factor ~1.3 is seen in CO gas for the atomic 
case and high ionisation, showing that the higher ionisation rate 
can also facilitate gas-phase formation as well increased ice processing.  

Moving outwards to 30~AU, where most volatiles have accreted onto 
grain surfaces, different behaviour is seen yet again.  
This time \ce{CH4} ice is enhanced by a factor of ~1.7 to 2.4 at
the expense of CO gas and ice (with the latter depleted by almost 
an order of magnitude) for the high ionisation cases. 
\ce{H2O} and \ce{CO2} ice are enhanced and depleted by 
up to 22\% and 17\% respectively, with high ionisation only.  
Little change is seen for low ionisation, despite beginning 
the calculation with atomic abundances.  
For this scenario, it appears that in the outer disk (30~AU), 
the resulting abundances of \ce{CO2} and \ce{H2O} ice reproduce the inherited 
values to within a few percent.  

Finally, it is important to recognize that at all radii and in all
scenarios, including the full reset case, the chemistry is not in
thermodynamic equilibrium, i.e., the abundances are not simply set by
the overall elemental abundances and pressure \citep[see also the
discussion in][]{henning13}.

\begin{figure*}
\subfigure{\includegraphics[width=0.5\linewidth]{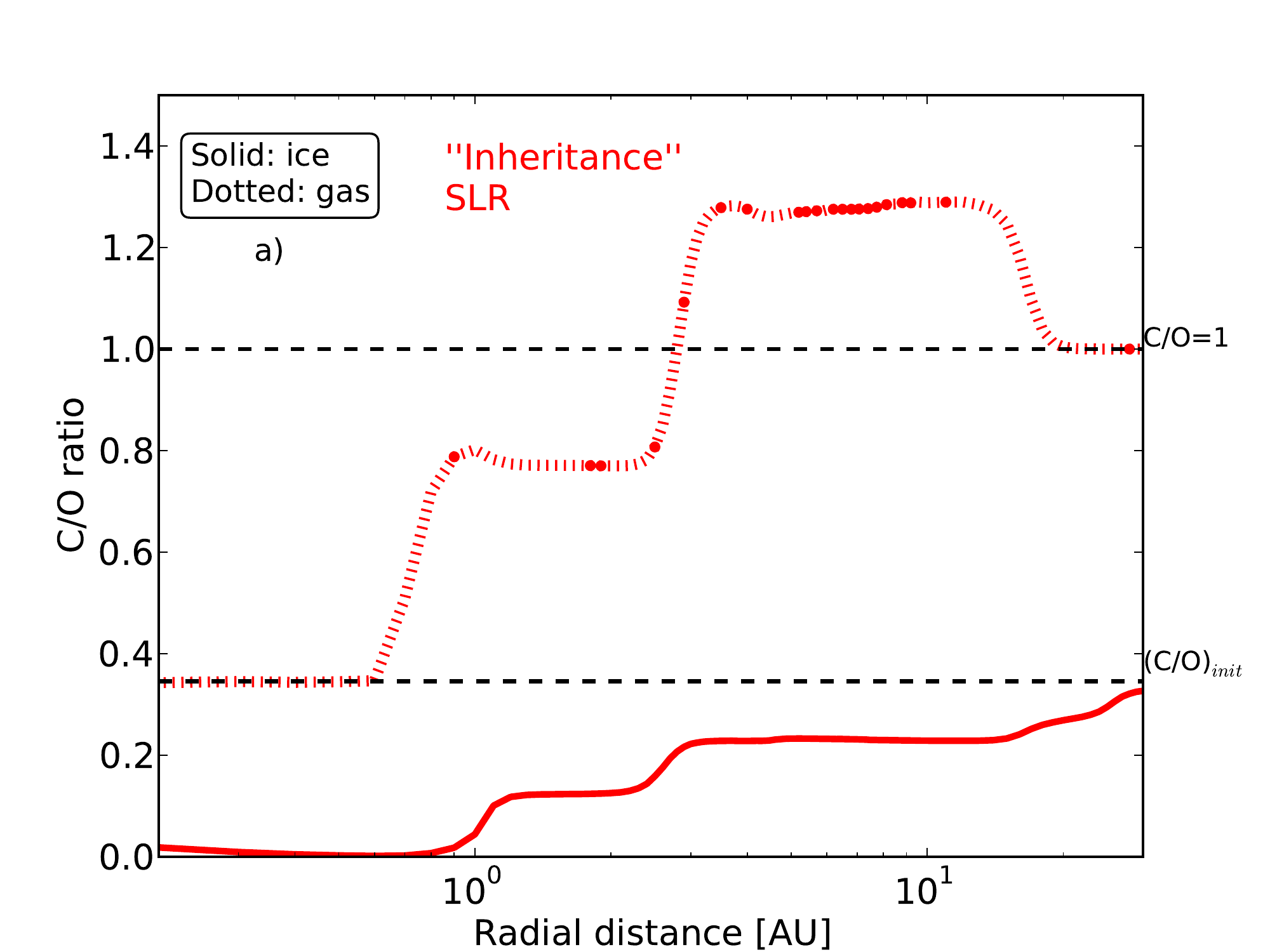}\label{c_o_mol_croff}}
\subfigure{\includegraphics[width=0.5\linewidth]{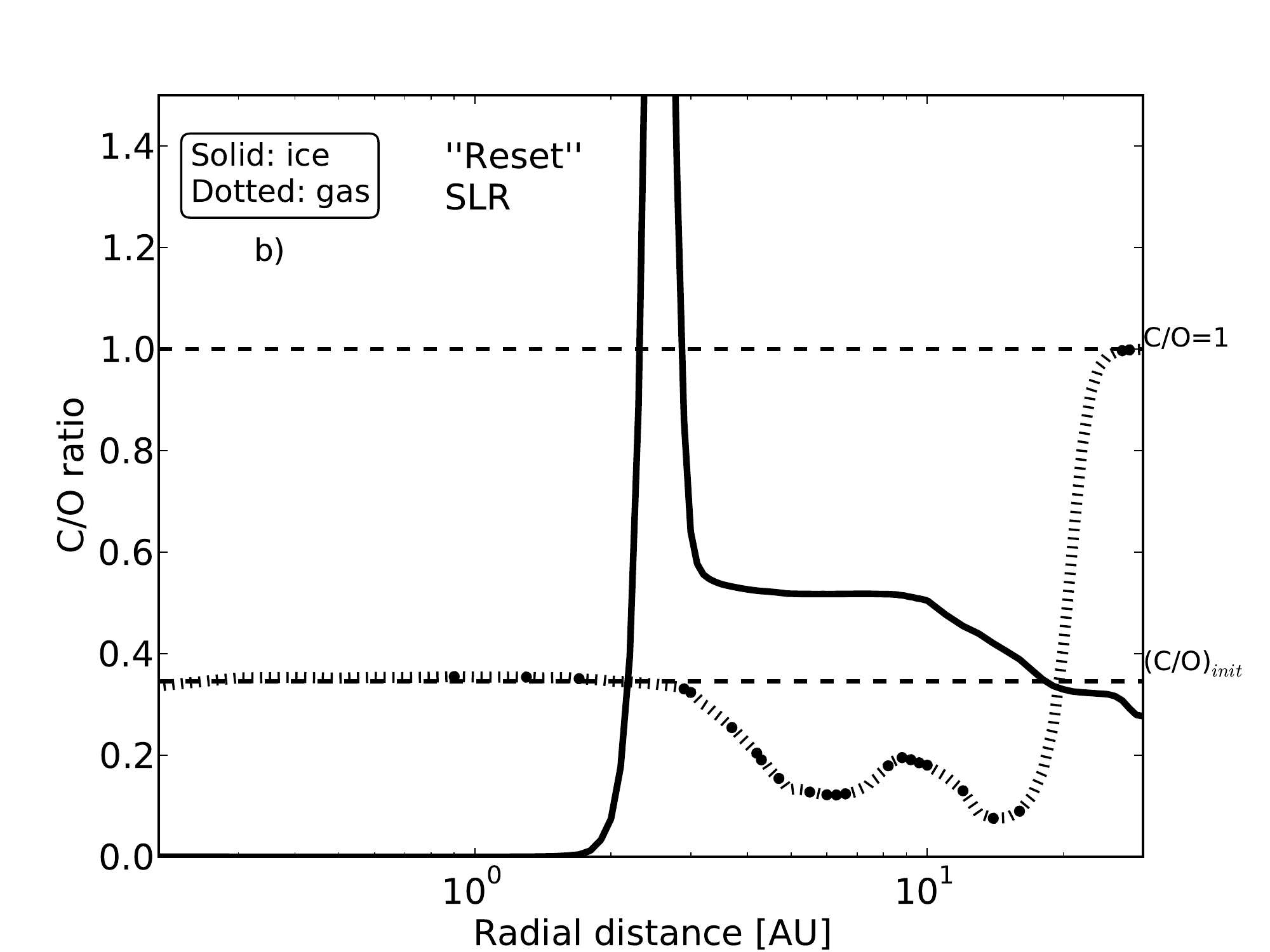}\label{c_o_atom_croff}}\\
\subfigure{\includegraphics[width=0.5\linewidth]{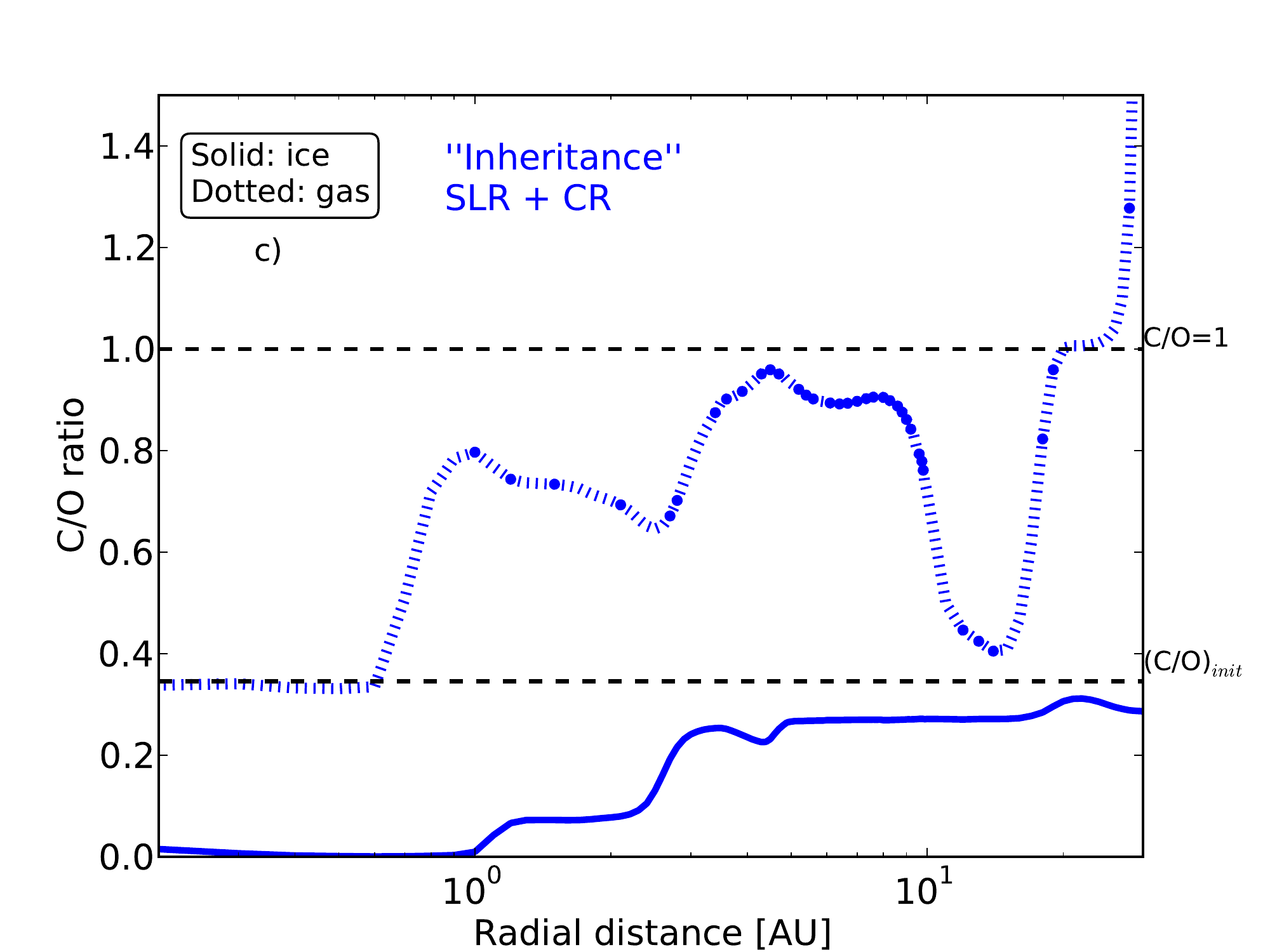}\label{c_o_mol_cron}}
\subfigure,{\includegraphics[width=0.5\linewidth]{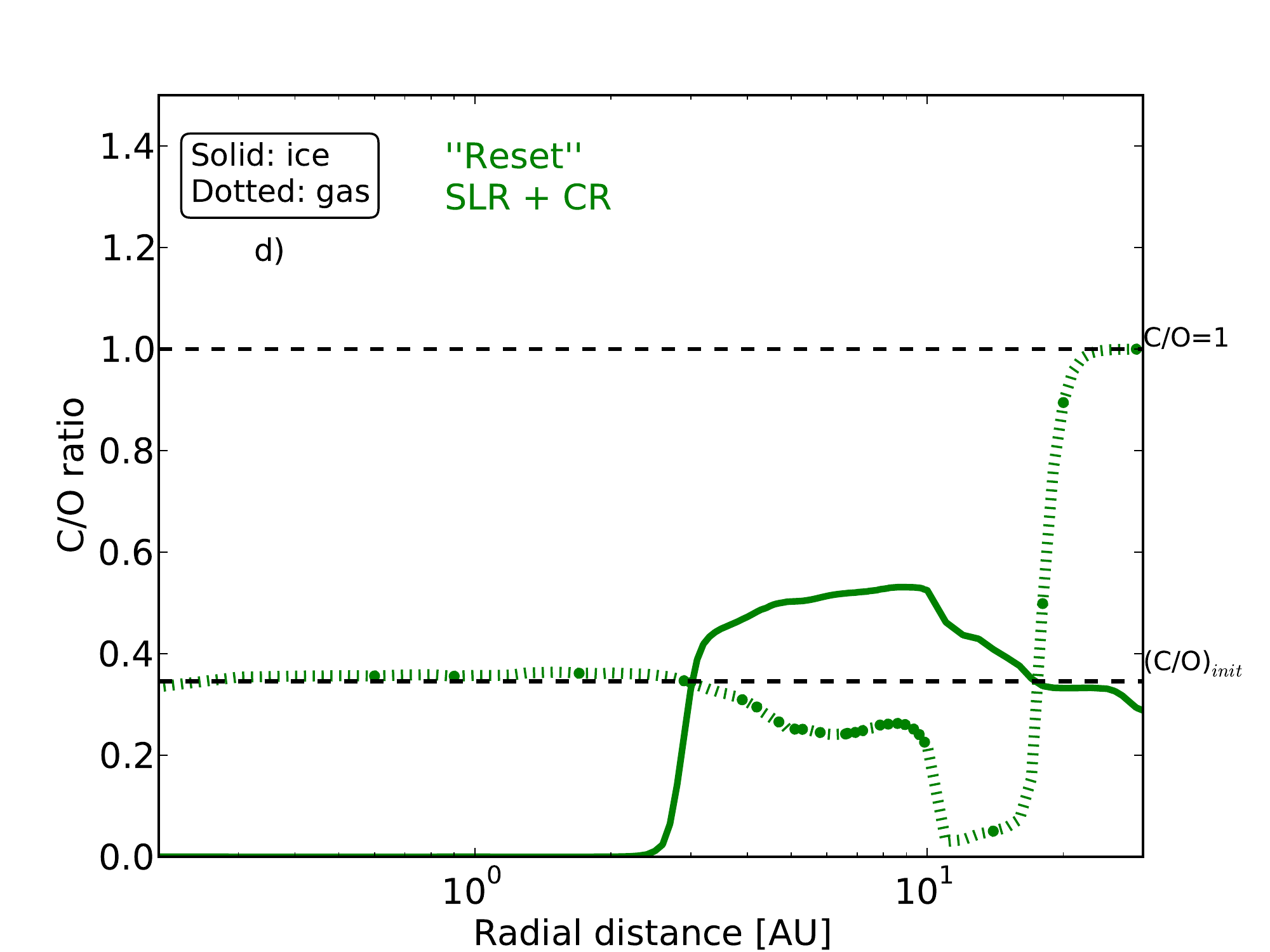}\label{c_o_atom_cron}}
\caption{C/O ratios in the gas (dashed lines) and ice (solid lines) for the 4 full chemistry model setups as described in the text. \textbf{The solid profile in panel b) peaks at C/O ratio 2.5}.}
\label{c_o}
\end{figure*}

\subsection{C/O ratio}

The results in Table \ref{result_composition} show that for different
sets of assumptions or models setups, a large diversity is seen in the
resulting computed abundances of dominant C- and O-bearing volatiles.
The deviations are also radially dependent, and align with the positions
of icelines.  The fractional deviations are sufficiently large to
affect the C/O ratio in the gas and ice which will go into forming the
building blocks of planets.  


Fig.~\ref{c_o} shows the C/O ratios for gas (dashed) and ice (solid)
species in the four different model setups using the full chemistry.
In the calculation of the ratio, all dominant C- and O-bearing species
were taken into account.  In addition to the main considered volatiles
(\ce{CO}, \ce{H2O}, \ce{CO2} and \ce{CH4}), these also include
\ce{CH3OH}, \ce{O2}, \ce{HCN}, \ce{NO}, and atomic O where
appropriate.  The horizontal lines indicate the canonical C/O ratio
(at 0.43) and C/O = 1.

For the inheritance scenario and low ionisation (Fig. \ref{c_o}a), the C/O ratio for the gas resembles a step
function.  Moving outwards in radius, the steps coincide with the
icelines of \ce{H2O}, \ce{CO2}, and \ce{CH4}, respectively.  This
profile is very similar to that in Fig.~4 in \citet{oberg2011co} in
which the C/O ratios in the gas and ice were assumed to be dictated
solely by the positions of icelines.  Only for the case of
inheritance and low ionisation level is there a region in the
disk, between 3 and 16 AU, in which the gas is carbon rich, i.e., C/O
$> 1$.  The ice ratio for the same model shows that overall the ice
remains carbon poor (or oxygen rich), but does become relatively more
carbon rich moving outwards in the disk as the icelines for \ce{CO2}
and \ce{CH4} are surpassed.

Considering the inheritance scenario with the high ionisation
level (Fig. \ref{c_o}c), the chemical processing
induced by CRs has a noticeable affect on the C/O ratio in the gas
and ice.  The C/O gas-phase ratio remains less than 1 everywhere due
to the destruction of \ce{CH4} gas and the production of \ce{O2} gas.
The gas-phase C/O ratio appears to increase significantly beyond
26~AU; however, this is beyond the CO iceline where most molecules
(except \ce{H2}) are depleted from the gas.  The C/O ice ratio looks
similar to that for the case of low ionisation.  The ratio is a bit
higher for the low ionisation case within 3~AU (73\% at 1.6 AU), and
slightly lower beyond 3~AU (15\% at 10 AU) reflecting the
repartitioning of atomic carbon and oxygen from \ce{H2O} ice into
\ce{CO2}, \ce{CO}, and \ce{O2} (see Fig.~\ref{full_chem_plots}).

Turning attention to the reset scenarios in Figs.~\ref{c_o}b and \ref{c_o}d, the picture changes significantly.  Between 3
and 16 AU the C/O ratio in the ice is higher than that in the gas
(although both ice and gas remain carbon poor).  This is opposite to
both inheritance cases; the dominant carriers of gas-phase C and O
in these models in this region are CO and \ce{O2}, and the dominant
ice component is \ce{CO2} ice as opposed to \ce{H2O} ice.  Thus the
C/O ice ratios tends towards $\approx 0.5$ whereas in the gas it tends
towards $\approx 0.3$.

The peak at $\approx 2.5$~AU for the profile in
Fig.~\ref{c_o}d is due to HCN ice being produced and freezing out
at a slightly higher temperature than \ce{CO2}.  This causes an
increase in the C/O ice ratio in this local region (see
Fig. \ref{nitros}b).  It is only seen in the low ionisation case
because N-bearing species formed via gas-phase chemistry in the inner
disk ($< 3$ AU) are able to survive to $10^{6}$~yrs.  The presence or
otherwise of this large peak is dependent upon the relative binding
energies of HCN and \ce{CO2} assumed in the model; recent
measurements of thermal desorption of pure HCN ice do derive a higher
binding energy than both \ce{CO2} and \ce{NH3}
\citep[e.g.,][]{noble2013}, but in a mixed ice they may be more similar.

The C/O ratio plots in Fig. \ref{c_o} highlight that the different
model setups considered here (especially the inheritance versus
reset scenarios) result in very different C/O ratio profiles for
the material in the planet-forming regions of the disk midplane. 

\subsection{Implications for planet formation and comets}
\label{planet_imp}

\subsubsection{Giant planet atmospheres}

Giant planets can accrete their atmospheres either directly from the
surrounding gas or through accretion of icy planetesimals, or both. In addition, radial migration of a forming planet can influence the makeup of the resulting atmosphere as the planet moves through a gradient in gas and icy planetesimal composition.
Fig.~\ref{c_o}a suggests that if only accretion from the gas is
considered, a gas-giant planet forming between 3 and 16~AU may be able
to form a carbon-rich atmosphere in the case of inheritance with
low ionisation rates. For high ionisation rates, the atmosphere would
still be relatively carbon rich with respect to the canonical ratio
(horizontal dashed lines at C/O ratio 0.34 in Figs.~\ref{c_o}), yet with a ratio which remains $<1$.
In both cases, the atmosphere can be polluted by accreting icy
planetesimals which are oxygen rich, lowering C/O in the planet's atmosphere.

For the reset scenario, a gas-giant planet forming between 3 and
16 AU would accrete C-poor gas; however, if the volatile component
accreted by the planet were dominated by icy planetesimals, the
resulting atmosphere may become carbon rich relative to the canonical
value.

The low abundance of ice (relative to the gas) inside the \ce{CO2}
iceline for the reset scenario is also interesting from the
perspective of the overall core-envelope partitioning.  If the core of
the planet was to form from the solid material available at 1~AU in
Fig.~\ref{full_chem_plots}d, then the bulk of the planet would not be
built up of volatile ices, but rather of more refractory components
(rocks).  The composition of the forming atmosphere will then be set
solely by the composition of gas accreted onto the forming planet.

The \ce{CO2}-rich ice mantles formed beyond 16 AU in the reset scenario also
have an interesting implication for the first stage of planet
formation, the formation of pebble-sized objects.  The sticking
efficiency of $100 \mu$m-sized \ce{CO2} ice particles was recently
determined to be an order of magnitude lower than that for similarly
sized \ce{H2O} ice particles \citep{musiolik2016}.  Hence, under these
particular conditions, the first steps of planet formation may be
impeded.

\subsubsection{Cometary composition}

With molecular initial abundances (i.e., inheritance), the
composition is generally preserved (with a few already mentioned
exceptions, see Sect.~\ref{results}).  Hence, planetesimals forming
under these specific conditions will be composed of ``inherited''
material.  The overlap in the composition of comets, considered
pristine remnants of the solar nebula, and interstellar ices certainly
supports the hypothesis that a significant fraction of disk midplane
material may be inherited from the molecular cloud \citep[see,
e.g.,][]{charnley11,oberg2011ices}. For the reset case, cometary
and interstellar abundances would be significantly different, at least
for comets formed inside 30 AU.

It is interesting to consider whether a protoplanetary disk midplane,
which inherits only water ice, is able to synthesis \ce{O2} at a level
similar to that seen in comets because this could be an observational
constraint on the inheritance scenario.  Our models can produce
significant amounts of O$_2$ gas in the inner disk, as was also seen
and discussed in full 2D protoplanetary disk models of
\citet{walsh15}.  \citet{bieler15} find an \ce{O2}-to-\ce{H2O} ice
ratio of $\approx 4$\% in comet~67P/C-G, with \ce{O2} 
closely associated with the \ce{H2O} ice. They concluded \ce{O2} was likely
primordial in origin, i.e., originating from the parent molecular
cloud and/or protostellar envelope.  This result is surprising due to
the high reactivity of \ce{O2} ice.  
The confirmation of \ce{O2} in comet 1P/Halley at a level similar to
that in 67P/C-G suggests that \ce{O2} is a dominant cometary component
\citep{rubin2015}.  In Figs.~\ref{full_chem_plots}a and
\ref{full_chem_plots}c, molecular oxygen is produced only for the high
ionisation case and reaches a peak abundance of $<0.1$\% that of water
ice.  This reflects the efficient conversion of \ce{O2} into water
ice, once formed.  The reset case with low ionisation produces
more O$_2$ gas and ice with respect to \ce{H2O} ice, but it remains
difficult to envisage a scenario in which H$_2$O and O$_2$ ice would
be well mixed.  Nevertheless, these results are sufficiently
interesting to warrant a detailed study exploring a broader range of
parameter space, which will be conducted in future work.

HCN has been detected in numerous comets with an abundance ratio of
the order of a few $\times 0.1$\% relative to water \citep[see,
e.g.,][]{charnley11,wirstrom2016}.  The results presented here show
that HCN ice is unlikely to be produced in-situ via the processing of ices
containing only \ce{N2} and \ce{NH3}, which is opposite to the case
for \ce{O2}.

\subsection{Caveats of model assumptions}
\label{caveats}

Several assumptions have been made in our models, in addition to the
set of initial abundances, the ionisation rate, and the chemical
network used. We have considered only a single temperature and
density profile, as well as single dust-grain size and density.
Although the disk surface density used is typical of those derived for
nearby protoplanetary disks \citep[see, e.g.,][]{williams11}, in
reality, disk midplane temperature profiles depend on a multitude of
parameters including, disk mass, stellar spectral type, dust
distribution and opacity, and disk flaring index (which influences the
amount of UV absorbed and/or scattered by the disk surface layer).
The iceline locations determined here are consistent with those
determined in protoplanetary disk models which are considered to be
representative of the pre-solar nebula \citep[see, e.g.,][and
references therein]{oberg2011co}.  For disks around warmer stars, the
icelines would move outwards in radius, with the opposite result for
disks around colder stars.  All else being equal, this will only
affect the radial extent over which we see each chemical effect.  A
significantly warmer inner disk will also allow gas-phase reactions
with reactions to occur more readily which may perturb the chemistry
differently to that seen here. Also, our adopted disk structure is static in time. This was chosen to keep focus on chemical evolution for fixed physical conditions. Future work will utilise an physical disk model that is evolving in time.

There is observational evidence for both dust-grain settling and growth 
in protoplanetary disks \citep[e.g.,][]{dullemond04,dullemond05,williams11}.  
Settling towards the midplane will increase the dust-to-gas mass ratio, whereas dust 
growth to $\approx$~mm sizes, will act in the opposite sense to decrease the total 
available surface area for freezeout and grain-surface reactions.  
The combination of both effects can either decrease or increase the total 
surface area of dust grains per unit volume.  
A smaller surface area (i.e., generally larger grains) will decrease the rate of 
freezeout (although freezeout is found to be almost instantaneous at the high densities of disk midplanes), 
and decrease the rate of grain-surface processing of ice mantles.  
Hence, this will lessen the effects of grain-surface chemistry, particularly 
for the case of high ionisation.  
Larger dust grains are also able to drift inwards as they become decoupled 
from the gas and affect the locations of icelines, by as much as 60\% 
\citep[as argued by][]{piso2015}. The effects of different grain sizes will be investigated and addressed in future work.
 
Only two ionisation levels have been assumed here, to probe what are considered to be two 
extreme cases; however, there is also speculation in the literature that 
the pre-solar nebula may have been exposed to an increased cosmic-ray rate due to nearby 
supernovae early in the Sun's lifetime \citep[see, e.g., the review by][]{adams2010}.  
We have demonstrated here that a relatively high level of ionisation causes more chemical 
processing than a lower level. 
We would expect this trend to continue for ionisation levels $\gg 10^{-17}$~s$^{-1}$.  
A lower level of ionisation would likely not have any effect, as the low level considered here 
is already seen not to significantly alter the chemistry. 
An increase or decrease in total disk mass (or surface density) will have the same 
effect as decreasing and increasing the ionisation rate due to CRs 
(especially in the outer disk), and increasing and decreasing the ionisation rate
due to SLRs (especially in the inner disk).  

The choice of initial abundances in the inheritance-scenario assumes an instant transition from molecular cloud phase to the disk midplane for the gas, with no chemical processing taking place ``on the way''. Likewise, the reset-scenario assumes all gas content to be dissociated into atoms at the time it reaches the midplane. These two sets of abundances are extremes. Recent models indicate that chemical processing of the material does take place on its way to the midplane (Drozdovskaya et al., MNRAS 2016, submitted), so a realistic initial composition of the material is more likely to be a mix of atoms and molecules. Considering the extreme cases, however, provides insight into the degree of chemical processing taking place.

Finally, we extracted our chemical abundances at a single time step only, at $10^{6}$~yrs, 
which is considered to be representative of the lifetime of the pre-solar nebula and nearby young 
protoplanetary disks \citep[e.g.,][]{williams11,henning13}.  
A shorter time would lead to less extreme effects due to ion-molecule reactions and CR-induced 
photoreactions, because these reactions typically have long timescales ($\gtrsim 10^{5}$~yrs) (see Fig.~\ref{meth_evol}); 
however, this strongly depends on the assumed cosmic-ray ionisation rate.   
A longer time would lead to the opposite case.
The ALMA detection of gaps and rings in the disk around the young ($<10^{6}$~yr) protostar, 
HL~Tau, suggests that grain growth and planet(esimal) formation in protoplanetary disks may occur much earlier than 
heretofore considered \citep{alma2015}; 
hence, shorter chemical timescales are worthy of further investigation in future work. Specifically, in our models, assuming the inheritance-scenario and high level of ionisation, significant changes to the abundances of key volatiles take place after a few times $10^{5}$ yrs (for more details, see Appendix \ref{timescale-app}).



\section{Conclusions}
\label{conclusions}

The models presented in this work have examined the importance 
of kinetic chemistry on the molecular composition 
(gas and ice) in protoplanetary disk midplanes.  
The main conclusions are listed below.

\begin{itemize}

\item The disk midplane composition reflects that of interstellar ices
  only for the case of low ionisation (SLRs only) in the inheritance
  scenario.  The partitioning between gas and ice is
  determined solely by iceline positions, as is assumed in the planet
  population synthesis models.  The inclusion of grain-surface
  chemistry has a negligible effect.

\item Assuming a higher rate of ionisation (SLRs plus CRs) and
  inheritance leads to an increase in the abundance of \ce{CH4} ice
  beyond its iceline, and a significant depletion of gas-phase
  \ce{CH4} in the critical region between 1 and
  15~AU. Cosmic-ray-induced chemistry enables the release of free
  carbon from CO in the outer disk ($>15$~AU) which is incorporated
  into gas-phase methane which freezes out. This naturally leads to
  low gas-phase CO abundances as is observed in some disks. On the
  other hand, cosmic-ray-induced chemistry efficiently destroys
  methane gas in the \ce{CH4}-poor region.  The conclusion holds
  whether grain-surface chemistry is included or not.
 
\item When grain-surface chemistry is considered, the \ce{CO2} ice to
  \ce{H2O} ice ratio is increased, with \ce{CO} and \ce{CH4} gas
  destroyed at the expense of an increase in \ce{CO2} ice.  The
  critical reactions are the photodissociation of \ce{H2O} ice to form
  \ce{OH} radicals within the ice mantle, which subsequently react
  with CO to form \ce{CO2} ice.  This reaction is able to proceed
  faster than H~+~OH recombination within $\approx 20$~AU because the
  very volatile H atoms are quickly lost to the gas phase for grains
  warmer than 20~K.  Beyond this radius, the reformation of water ice
  wins.  The partitioning between \ce{N2} and \ce{NH3} is similarly
  affected.

\item For the extreme reset scenario in which all elements are
  initially in atomic form, the picture changes significantly.
  Without grain-surface chemistry, gas-phase CO, \ce{O2}, and atomic
  oxygen are the main carbon and oxygen-bearing species beyond 1~AU.
  The chemistry does not have sufficient time to incorporate all
  available initial elemental oxygen into molecules by $10^6$~yrs.
  Gas-phase water and \ce{CO2} do form and subsequently freeze out,
  albeit achieving much lower abundances than in the inheritance
  scenario.  A higher ionisation level helps to increase the
  production of \ce{H2O} and \ce{CH4}.

\item With grain-surface chemistry, the abundances of \ce{H2O} and
  \ce{CO2} ice increase significantly, demonstrating the absolute
  necessity of grain-surface chemistry for the synthesis of these two
  dominant ice components.  The final abundance ratios reached in the
  very outer disk (30~AU) for \ce{H2O} and \ce{CO2} are similar to
  those for the inheritance scenario, regardless of the ionisation
  level; however, the higher ionisation level does impede the
  abundance of \ce{O2} ice at the levels seen in comets and enables a
  conversion from CO to \ce{CH4}.

\item In the reset scenario, species other than the main
  considered volatiles are also produced in non-negligible quantities:
  \ce{O2}, \ce{HCN}, and \ce{NO}.  The higher ionisation level
  generally helps the production of HCN and NO (at the expense of
  \ce{N2} and \ce{NH3}) and impedes the survival of \ce{O2} ice.

\item The inclusion of chemistry has a significant impact on the C/O
  ratio of both gas and ice in the planet-forming region which is
  expected to influence the resulting composition of forming
  planet(esimal)s.  The ices remain, on the whole, dominated
  by oxygen (i.e., C/O $< 1$).  For both inheritance cases, the gas is
  carbon rich relative to the canonical value; however only for the
  low ionisation case is there a reservoir of gas-phase material with
  C/O $>1$.  For both reset scenarios, the ice becomes more
  carbon rich than the gas, which is opposite to the inheritance
  case.

\end{itemize}

The results presented here show that under certain conditions,
highlighted above, chemistry can have a profound effect on the
composition of the planet-forming material in disk midplanes.
Chemistry influences the partitioning of elemental carbon, oxygen, and
nitrogen, into molecules of differing volatilities, such that the
positions of ice lines alone, are not necessarily adequate for
determining the ratio of C/O in neither the gas, nor the ice.  Only
under the extreme case of full inheritance and low ionisation, are the
elemental ratios determined solely by the positions of icelines.  This
conclusion is also similar to that for the assumption that ices are
already locked up in larger bodies by $\approx 10^{5}$~yr.

The work presented here follows the time evolution of chemistry in a
static protoplanetary disk which is the simplest physical case.  In
reality, disk conditions evolve with time, at the same time as
planetesimals are forming and migrating within the midplane.  Future
plans include determining the influence of chemistry in an evolving
protoplanetary disk (where the density, temperature, and ionisation
rate also vary with time), and to couple the outputs of these models
with planet formation tracks to determine, in a quantitative manner,
the influence on the resulting composition of gas-giant planetary
atmospheres.

\begin{acknowledgements}
The authors thank Amaury Thiabaud and Ulysses Marboeuf for making
their disk density and temperature profile available and for many
inspiring discussions. The authors also thank Karin \"Oberg for a fruitful discussion about CO gas depletion inside the CO iceline. Astrochemistry in Leiden is supported by the
European Union A-ERC grant 291141 CHEMPLAN, by the Netherlands
Research School for Astronomy (NOVA), and by a Royal Netherlands
Academy of Arts and Sciences (KNAW) professor prize. CW also
acknowledges the Netherlands Organisation for Scientific Research
(NWO, grant 639.041.335).
\end{acknowledgements}

\bibliographystyle{aa} 
\bibliography{bib_new} 

\begin{thebibliography}{83}
\expandafter\ifx\csname natexlab\endcsname\relax\def\natexlab#1{#1}\fi

\bibitem[{{Adams}(2010)}]{adams2010}
{Adams}, F.~C. 2010, \araa, 48, 47

\bibitem[{{Aikawa} \& {Herbst}(1999)}]{aikawa99}
{Aikawa}, Y. \& {Herbst}, E. 1999, \aap, 351, 233

\bibitem[{{Aikawa} {et~al.}(1999){Aikawa}, {Umebayashi}, {Nakano}, \&
  {Miyama}}]{aikawa1999accre}
{Aikawa}, Y., {Umebayashi}, T., {Nakano}, T., \& {Miyama}, S.~M. 1999, \apj,
  519, 705

\bibitem[{{Alibert} {et~al.}(2013){Alibert}, {Carron}, {Fortier}, {Pfyffer},
  {Benz}, {Mordasini}, \& {Swoboda}}]{alibert13}
{Alibert}, Y., {Carron}, F., {Fortier}, A., {et~al.} 2013, \aap, 558, A109

\bibitem[{{ALMA Partnership} {et~al.}(2015){ALMA Partnership}, {Brogan},
  {P{\'e}rez}, {Hunter}, {Dent}, {Hales}, {Hills}, {Corder}, {Fomalont},
  {Vlahakis}, {Asaki}, {Barkats}, {Hirota}, {Hodge}, {Impellizzeri}, {Kneissl},
  {Liuzzo}, {Lucas}, {Marcelino}, {Matsushita}, {Nakanishi}, {Phillips},
  {Richards}, {Toledo}, {Aladro}, {Broguiere}, {Cortes}, {Cortes}, {Espada},
  {Galarza}, {Garcia-Appadoo}, {Guzman-Ramirez}, {Humphreys}, {Jung}, {Kameno},
  {Laing}, {Leon}, {Marconi}, {Mignano}, {Nikolic}, {Nyman}, {Radiszcz},
  {Remijan}, {Rod{\'o}n}, {Sawada}, {Takahashi}, {Tilanus}, {Vila Vilaro},
  {Watson}, {Wiklind}, {Akiyama}, {Chapillon}, {de Gregorio-Monsalvo}, {Di
  Francesco}, {Gueth}, {Kawamura}, {Lee}, {Nguyen Luong}, {Mangum}, {Pietu},
  {Sanhueza}, {Saigo}, {Takakuwa}, {Ubach}, {van Kempen}, {Wootten},
  {Castro-Carrizo}, {Francke}, {Gallardo}, {Garcia}, {Gonzalez}, {Hill},
  {Kaminski}, {Kurono}, {Liu}, {Lopez}, {Morales}, {Plarre}, {Schieven},
  {Testi}, {Videla}, {Villard}, {Andreani}, {Hibbard}, \&
  {Tatematsu}}]{alma2015}
{ALMA Partnership}, {Brogan}, C.~L., {P{\'e}rez}, L.~M., {et~al.} 2015, \apjl,
  808, L3

\bibitem[{{Andrews} {et~al.}(2010){Andrews}, {Wilner}, {Hughes}, {Qi}, \&
  {Dullemond}}]{andrews10}
{Andrews}, S.~M., {Wilner}, D.~J., {Hughes}, A.~M., {Qi}, C., \& {Dullemond},
  C.~P. 2010, \apj, 723, 1241

\bibitem[{{Armitage}(2011)}]{armitage2011}
{Armitage}, P.~J. 2011, \araa, 49, 195

\bibitem[{{Batalha} {et~al.}(2013){Batalha}, {Rowe}, {Bryson}, {Barclay},
  {Burke}, {Caldwell}, {Christiansen}, {Mullally}, {Thompson}, {Brown},
  {Dupree}, {Fabrycky}, {Ford}, {Fortney}, {Gilliland}, {Isaacson}, {Latham},
  {Marcy}, {Quinn}, {Ragozzine}, {Shporer}, {Borucki}, {Ciardi}, {Gautier},
  {Haas}, {Jenkins}, {Koch}, {Lissauer}, {Rapin}, {Basri}, {Boss}, {Buchhave},
  {Carter}, {Charbonneau}, {Christensen-Dalsgaard}, {Clarke}, {Cochran},
  {Demory}, {Desert}, {Devore}, {Doyle}, {Esquerdo}, {Everett}, {Fressin},
  {Geary}, {Girouard}, {Gould}, {Hall}, {Holman}, {Howard}, {Howell},
  {Ibrahim}, {Kinemuchi}, {Kjeldsen}, {Klaus}, {Li}, {Lucas}, {Meibom},
  {Morris}, {Pr{\v s}a}, {Quintana}, {Sanderfer}, {Sasselov}, {Seader},
  {Smith}, {Steffen}, {Still}, {Stumpe}, {Tarter}, {Tenenbaum}, {Torres},
  {Twicken}, {Uddin}, {Van Cleve}, {Walkowicz}, \& {Welsh}}]{batalha2013}
{Batalha}, N.~M., {Rowe}, J.~F., {Bryson}, S.~T., {et~al.} 2013, \apjs, 204, 24

\bibitem[{{Benz} {et~al.}(2014){Benz}, {Ida}, {Alibert}, {Lin}, \&
  {Mordasini}}]{benz2014}
{Benz}, W., {Ida}, S., {Alibert}, Y., {Lin}, D., \& {Mordasini}, C. 2014, in
  Protostars and Planets VI, ed. H.~{Beuther}, R.~F. {Klessen}, C.~P.
  {Dullemond}, \& T.~{Henning} ({The University of Arizona Press}), 691--713

\bibitem[{{Bergin} {et~al.}(2007){Bergin}, {Aikawa}, {Blake}, \& {van
  Dishoeck}}]{bergin2007}
{Bergin}, E.~A., {Aikawa}, Y., {Blake}, G.~A., \& {van Dishoeck}, E.~F. 2007,
  in Protostars and Planets V, ed. B.~{Reipurth}, D.~{Jewitt}, \& K.~{Keil}
  ({The University of Arizona Press}), 751--766

\bibitem[{Bieler {et~al.}(2015)Bieler, Altwegg, Balsiger, Bar-Nun, Berthelier,
  Bochsler, Briois, Calmonte, Combi, De~Keyser, van Dishoeck, Fiethe, Fuselier,
  Gasc, Gombosi, Hansen, Hassig, Jackel, Kopp, Korth, Le~Roy, Mall, Maggiolo,
  Marty, Mousis, Owen, Reme, Rubin, Semon, Tzou, Waite, Walsh, \&
  Wurz}]{bieler15}
Bieler, A., Altwegg, K., Balsiger, H., {et~al.} 2015, Nature, 526, 678

\bibitem[{{Birkby} {et~al.}(2013){Birkby}, {de Kok}, {Brogi}, {de Mooij},
  {Schwarz}, {Albrecht}, \& {Snellen}}]{birkby2013}
{Birkby}, J.~L., {de Kok}, R.~J., {Brogi}, M., {et~al.} 2013, \mnras, 436, L35

\bibitem[{{Boogert} {et~al.}(2015){Boogert}, {Gerakines}, \&
  {Whittet}}]{boogert2015}
{Boogert}, A.~C.~A., {Gerakines}, P.~A., \& {Whittet}, D.~C.~B. 2015, \araa,
  53, 541

\bibitem[{{Borucki} {et~al.}(2011){Borucki}, {Koch}, {Basri}, {Batalha},
  {Brown}, {Bryson}, {Caldwell}, {Christensen-Dalsgaard}, {Cochran}, {DeVore},
  {Dunham}, {Gautier}, {Geary}, {Gilliland}, {Gould}, {Howell}, {Jenkins},
  {Latham}, {Lissauer}, {Marcy}, {Rowe}, {Sasselov}, {Boss}, {Charbonneau},
  {Ciardi}, {Doyle}, {Dupree}, {Ford}, {Fortney}, {Holman}, {Seager},
  {Steffen}, {Tarter}, {Welsh}, {Allen}, {Buchhave}, {Christiansen}, {Clarke},
  {Das}, {D{\'e}sert}, {Endl}, {Fabrycky}, {Fressin}, {Haas}, {Horch},
  {Howard}, {Isaacson}, {Kjeldsen}, {Kolodziejczak}, {Kulesa}, {Li}, {Lucas},
  {Machalek}, {McCarthy}, {MacQueen}, {Meibom}, {Miquel}, {Prsa}, {Quinn},
  {Quintana}, {Ragozzine}, {Sherry}, {Shporer}, {Tenenbaum}, {Torres},
  {Twicken}, {Van Cleve}, {Walkowicz}, {Witteborn}, \& {Still}}]{borucki2011}
{Borucki}, W.~J., {Koch}, D.~G., {Basri}, G., {et~al.} 2011, \apj, 736, 19

\bibitem[{{Bruderer}(2013)}]{bruderer2013}
{Bruderer}, S. 2013, \aap, 559, A46

\bibitem[{{Bruderer} {et~al.}(2014){Bruderer}, {van der Marel}, {van Dishoeck},
  \& {van Kempen}}]{bruderer14}
{Bruderer}, S., {van der Marel}, N., {van Dishoeck}, E.~F., \& {van Kempen},
  T.~A. 2014, \aap, 562, A26

\bibitem[{{Ceccarelli} {et~al.}(2014){Ceccarelli}, {Caselli},
  {Bockel{\'e}e-Morvan}, {Mousis}, {Pizzarello}, {Robert}, \&
  {Semenov}}]{ceccarelli2014}
{Ceccarelli}, C., {Caselli}, P., {Bockel{\'e}e-Morvan}, D., {et~al.} 2014,
  Protostars and Planets VI, 859

\bibitem[{{Cleeves} {et~al.}(2013{\natexlab{a}}){Cleeves}, {Adams}, \&
  {Bergin}}]{cleeves13crex}
{Cleeves}, L.~I., {Adams}, F.~C., \& {Bergin}, E.~A. 2013{\natexlab{a}}, \apj,
  772, 5

\bibitem[{{Cleeves} {et~al.}(2013{\natexlab{b}}){Cleeves}, {Adams}, {Bergin},
  \& {Visser}}]{cleeves13slr}
{Cleeves}, L.~I., {Adams}, F.~C., {Bergin}, E.~A., \& {Visser}, R.
  2013{\natexlab{b}}, \apj, 777, 28

\bibitem[{{Cleeves} {et~al.}(2014{\natexlab{a}}){Cleeves}, {Bergin}, \&
  {Adams}}]{cleeves14crex}
{Cleeves}, L.~I., {Bergin}, E.~A., \& {Adams}, F.~C. 2014{\natexlab{a}}, \apj,
  794, 123

\bibitem[{{Cleeves} {et~al.}(2014{\natexlab{b}}){Cleeves}, {Bergin},
  {Alexander}, {Du}, {Graninger}, {{\"O}berg}, \& {Harries}}]{cleeves2014water}
{Cleeves}, L.~I., {Bergin}, E.~A., {Alexander}, C.~M.~O., {et~al.}
  2014{\natexlab{b}}, Science, 345, 1590

\bibitem[{{Crossfield}(2015)}]{crossfield2015}
{Crossfield}, I.~J.~M. 2015, \pasp, 127, 941

\bibitem[{{Du} {et~al.}(2015){Du}, {Bergin}, \& {Hogerheijde}}]{du2015}
{Du}, F., {Bergin}, E.~A., \& {Hogerheijde}, M.~R. 2015, \apjl, 807, L32

\bibitem[{{Dullemond} \& {Dominik}(2004)}]{dullemond04}
{Dullemond}, C.~P. \& {Dominik}, C. 2004, \aap, 421, 1075

\bibitem[{{Dullemond} \& {Dominik}(2005)}]{dullemond05}
{Dullemond}, C.~P. \& {Dominik}, C. 2005, \aap, 434, 971

\bibitem[{{Dullemond} {et~al.}(2007){Dullemond}, {Hollenbach}, {Kamp}, \&
  {D'Alessio}}]{dullemond2007}
{Dullemond}, C.~P., {Hollenbach}, D., {Kamp}, I., \& {D'Alessio}, P. 2007, in
  Protostars and Planets V, ed. B.~{Reipurth}, D.~{Jewitt}, \& K.~{Keil} ({The
  University of Arizona Press}), 555--572

\bibitem[{{Fedele} {et~al.}(2011){Fedele}, {Pascucci}, {Brittain}, {Kamp},
  {Woitke}, {Williams}, {Dent}, \& {Thi}}]{fedele2011}
{Fedele}, D., {Pascucci}, I., {Brittain}, S., {et~al.} 2011, \apj, 732, 106

\bibitem[{{Fischer} {et~al.}(2014){Fischer}, {Howard}, {Laughlin}, {Macintosh},
  {Mahadevan}, {Sahlmann}, \& {Yee}}]{fischer2014}
{Fischer}, D.~A., {Howard}, A.~W., {Laughlin}, G.~P., {et~al.} 2014, in
  Protostars and Planets VI, ed. H.~{Beuther}, R.~F. {Klessen}, C.~P.
  {Dullemond}, \& T.~{Henning} ({The University of Arizona Press}), 715--737

\bibitem[{{Fraine} {et~al.}(2014){Fraine}, {Deming}, {Benneke}, {Knutson},
  {Jord{\'a}n}, {Espinoza}, {Madhusudhan}, {Wilkins}, \&
  {Todorov}}]{fraine2014}
{Fraine}, J., {Deming}, D., {Benneke}, B., {et~al.} 2014, \nat, 513, 526

\bibitem[{{Furuya} \& {Aikawa}(2014)}]{furuya2014}
{Furuya}, K. \& {Aikawa}, Y. 2014, \apj, 790, 97

\bibitem[{{Furuya} {et~al.}(2016){Furuya}, {Drozdovskaya}, \&
  {Visser}}]{furuya2016}
{Furuya}, K., {Drozdovskaya}, M.~N., \& {Visser}, R. 2016, \aap, in press

\bibitem[{{Garrod} {et~al.}(2008){Garrod}, {Widicus Weaver}, \&
  {Herbst}}]{garrod2008}
{Garrod}, R.~T., {Widicus Weaver}, S.~L., \& {Herbst}, E. 2008, \apj, 682, 283

\bibitem[{{Gibb} {et~al.}(2004){Gibb}, {Whittet}, {Boogert}, \&
  {Tielens}}]{gibb2004}
{Gibb}, E.~L., {Whittet}, D.~C.~B., {Boogert}, A.~C.~A., \& {Tielens},
  A.~G.~G.~M. 2004, \apjs, 151, 35

\bibitem[{{Grossman}(1972)}]{grossman1972}
{Grossman}, L. 1972, \gca, 36, 597

\bibitem[{{Hasegawa} \& {Herbst}(1993)}]{hasegawa1993}
{Hasegawa}, T.~I. \& {Herbst}, E. 1993, \mnras, 261, 83

\bibitem[{{Henning} \& {Semenov}(2013)}]{henning13}
{Henning}, T. \& {Semenov}, D. 2013, Chemical Reviews, 113, 9016

\bibitem[{{Hollenbach} {et~al.}(2009){Hollenbach}, {Kaufman}, {Bergin}, \&
  {Melnick}}]{hollenbach2009}
{Hollenbach}, D., {Kaufman}, M.~J., {Bergin}, E.~A., \& {Melnick}, G.~J. 2009,
  \apj, 690, 1497

\bibitem[{{Ida} \& {Lin}(2004)}]{ida2004}
{Ida}, S. \& {Lin}, D.~N.~C. 2004, \apj, 604, 388

\bibitem[{{Ida} \& {Lin}(2008)}]{ida2008}
{Ida}, S. \& {Lin}, D.~N.~C. 2008, \apj, 673, 487

\bibitem[{{Johansen} {et~al.}(2014){Johansen}, {Blum}, {Tanaka}, {Ormel},
  {Bizzarro}, \& {Rickman}}]{johansen2014}
{Johansen}, A., {Blum}, J., {Tanaka}, H., {et~al.} 2014, in Protostars and
  Planets VI, ed. H.~{Beuther}, R.~F. {Klessen}, C.~P. {Dullemond}, \&
  T.~{Henning} ({The University of Arizona Press}), 547--570

\bibitem[{{Johnson} {et~al.}(2012){Johnson}, {Mousis}, {Lunine}, \&
  {Madhusudhan}}]{johnson2012}
{Johnson}, T.~V., {Mousis}, O., {Lunine}, J.~I., \& {Madhusudhan}, N. 2012,
  \apj, 757, 192

\bibitem[{{Kama} {et~al.}(2016{\natexlab{a}}){Kama}, {Bruderer}, {Carney},
  {Hogerheijde}, {van Dishoeck}, {Fedele}, {Baryshev}, {Boland}, {G{\"u}sten},
  {Aikutalp}, {Choi}, {Endo}, {Frieswijk}, {Karska}, {Klaassen}, {Koumpia},
  {Kristensen}, {Leurini}, {Nagy}, {Perez Beaupuits}, {Risacher}, {van der
  Marel}, {van Kempen}, {van Weeren}, {Wyrowski}, \&
  {Y{\i}ld{\i}z}}]{kama2016codepl}
{Kama}, M., {Bruderer}, S., {Carney}, M., {et~al.} 2016{\natexlab{a}}, \aap,
  588, A108

\bibitem[{{Kama} {et~al.}(2016{\natexlab{b}}){Kama}, {Bruderer}, {van
  Dishoeck}, {Hogerheijde}, {Folsom}, {Miotello}, {Fedele}, {Belloche},
  {G{\"u}sten}, \& {Wyrowski}}]{kama2016model}
{Kama}, M., {Bruderer}, S., {van Dishoeck}, E.~F., {et~al.} 2016{\natexlab{b}},
  ArXiv e-prints

\bibitem[{Linnartz {et~al.}(2015)Linnartz, Ioppolo, \& Fedoseev}]{linnartz2015}
Linnartz, H., Ioppolo, S., \& Fedoseev, G. 2015, International Reviews in
  Physical Chemistry, 34, 205

\bibitem[{{Marboeuf} {et~al.}(2014){Marboeuf}, {Thiabaud}, {Alibert}, {Cabral},
  \& {Benz}}]{marboeuf14}
{Marboeuf}, U., {Thiabaud}, A., {Alibert}, Y., {Cabral}, N., \& {Benz}, W.
  2014, \aap, 570, A35

\bibitem[{{McElroy} {et~al.}(2013){McElroy}, {Walsh}, {Markwick}, {Cordiner},
  {Smith}, \& {Millar}}]{mcelroy13}
{McElroy}, D., {Walsh}, C., {Markwick}, A.~J., {et~al.} 2013, \aap, 550, A36

\bibitem[{{Moses} {et~al.}(2013){Moses}, {Madhusudhan}, {Visscher}, \&
  {Freedman}}]{moses2013}
{Moses}, J.~I., {Madhusudhan}, N., {Visscher}, C., \& {Freedman}, R.~S. 2013,
  \apj, 763, 25

\bibitem[{{Mousis} {et~al.}(2010){Mousis}, {Lunine}, {Picaud}, \&
  {Cordier}}]{mousis2010}
{Mousis}, O., {Lunine}, J.~I., {Picaud}, S., \& {Cordier}, D. 2010, Faraday
  Discussions, 147, 509

\bibitem[{{Mumma} \& {Charnley}(2011)}]{charnley11}
{Mumma}, M.~J. \& {Charnley}, S.~B. 2011, \araa, 49, 471

\bibitem[{{Musiolik} {et~al.}(2016){Musiolik}, {Teiser}, {Jankowski}, \&
  {Wurm}}]{musiolik2016}
{Musiolik}, G., {Teiser}, J., {Jankowski}, T., \& {Wurm}, G. 2016, \apj, 818,
  16

\bibitem[{{Noble} {et~al.}(2013){Noble}, {Theule}, {Borget}, {Danger},
  {Chomat}, {Duvernay}, {Mispelaer}, \& {Chiavassa}}]{noble2013}
{Noble}, J.~A., {Theule}, P., {Borget}, F., {et~al.} 2013, \mnras, 428, 3262

\bibitem[{{Nomura} \& {Millar}(2005)}]{nomura2005}
{Nomura}, H. \& {Millar}, T.~J. 2005, \aap, 438, 923

\bibitem[{{Nomura} {et~al.}(2016){Nomura}, {Tsukagoshi}, {Kawabe}, {Ishimoto},
  {Okuzumi}, {Muto}, {Kanagawa}, {Ida}, {Walsh}, {Millar}, \&
  {Bai}}]{nomura2016}
{Nomura}, H., {Tsukagoshi}, T., {Kawabe}, R., {et~al.} 2016, \apjl, 819, L7

\bibitem[{{{\"O}berg} {et~al.}(2011{\natexlab{a}}){{\"O}berg}, {Boogert},
  {Pontoppidan}, {van den Broek}, {van Dishoeck}, {Bottinelli}, {Blake}, \&
  {Evans}}]{oberg2011ices}
{{\"O}berg}, K.~I., {Boogert}, A.~C.~A., {Pontoppidan}, K.~M., {et~al.}
  2011{\natexlab{a}}, \apj, 740, 109

\bibitem[{{{\"O}berg} {et~al.}(2011{\natexlab{b}}){{\"O}berg}, {Murray-Clay},
  \& {Bergin}}]{oberg2011co}
{{\"O}berg}, K.~I., {Murray-Clay}, R., \& {Bergin}, E.~A. 2011{\natexlab{b}},
  \apjl, 743, L16

\bibitem[{{Piso} {et~al.}(2015){Piso}, {{\"O}berg}, {Birnstiel}, \&
  {Murray-Clay}}]{piso2015}
{Piso}, A.-M.~A., {{\"O}berg}, K.~I., {Birnstiel}, T., \& {Murray-Clay}, R.~A.
  2015, \apj, 815, 109

\bibitem[{{Pontoppidan} {et~al.}(2014){Pontoppidan}, {Salyk}, {Bergin},
  {Brittain}, {Marty}, {Mousis}, \& {{\"O}berg}}]{pontoppidan2014}
{Pontoppidan}, K.~M., {Salyk}, C., {Bergin}, E.~A., {et~al.} 2014, in
  Protostars and Planets VI, ed. H.~{Beuther}, R.~F. {Klessen}, C.~P.
  {Dullemond}, \& T.~{Henning} ({The University of Arizona Press}), 363--385

\bibitem[{{Prasad} \& {Tarafdar}(1983)}]{prasad1983}
{Prasad}, S.~S. \& {Tarafdar}, S.~P. 1983, \apj, 267, 603

\bibitem[{{Reboussin} {et~al.}(2015){Reboussin}, {Wakelam}, {Guilloteau},
  {Hersant}, \& {Dutrey}}]{reboussin2015}
{Reboussin}, L., {Wakelam}, V., {Guilloteau}, S., {Hersant}, F., \& {Dutrey},
  A. 2015, \aap, 579, A82

\bibitem[{{Rubin} {et~al.}(2015){Rubin}, {Altwegg}, {van Dishoeck}, \&
  {Schwehm}}]{rubin2015}
{Rubin}, M., {Altwegg}, K., {van Dishoeck}, E.~F., \& {Schwehm}, G. 2015,
  \apjl, 815, L11

\bibitem[{{Schwarz} \& {Bergin}(2014)}]{schwarz2014}
{Schwarz}, K.~R. \& {Bergin}, E.~A. 2014, \apj, 797, 113

\bibitem[{{Schwarz} {et~al.}(2016){Schwarz}, {Bergin}, {Cleeves}, {Blake},
  {Zhang}, {{\"O}berg}, {van Dishoeck}, \& {Qi}}]{schwarz2016}
{Schwarz}, K.~R., {Bergin}, E.~A., {Cleeves}, L.~I., {et~al.} 2016, \apj, 823,
  91

\bibitem[{{Seager} \& {Deming}(2010)}]{seager2010}
{Seager}, S. \& {Deming}, D. 2010, \araa, 48, 631

\bibitem[{{Semenov} {et~al.}(2004){Semenov}, {Wiebe}, \&
  {Henning}}]{semenov2004}
{Semenov}, D., {Wiebe}, D., \& {Henning}, T. 2004, \aap, 417, 93

\bibitem[{{Sing} {et~al.}(2016){Sing}, {Fortney}, {Nikolov}, {Wakeford},
  {Kataria}, {Evans}, {Aigrain}, {Ballester}, {Burrows}, {Deming},
  {D{\'e}sert}, {Gibson}, {Henry}, {Huitson}, {Knutson}, {Etangs}, {Pont},
  {Showman}, {Vidal-Madjar}, {Williamson}, \& {Wilson}}]{sing2016}
{Sing}, D.~K., {Fortney}, J.~J., {Nikolov}, N., {et~al.} 2016, \nat, 529, 59

\bibitem[{{Snellen} {et~al.}(2010){Snellen}, {de Kok}, {de Mooij}, \&
  {Albrecht}}]{snellen2010}
{Snellen}, I.~A.~G., {de Kok}, R.~J., {de Mooij}, E.~J.~W., \& {Albrecht}, S.
  2010, \nat, 465, 1049

\bibitem[{{Thiabaud} {et~al.}(2015){Thiabaud}, {Marboeuf}, {Alibert}, {Leya},
  \& {Mezger}}]{thiabaud2015gascomp}
{Thiabaud}, A., {Marboeuf}, U., {Alibert}, Y., {Leya}, I., \& {Mezger}, K.
  2015, \aap, 574, A138

\bibitem[{{Tielens} \& {Hagen}(1982)}]{tielens1982}
{Tielens}, A.~G.~G.~M. \& {Hagen}, W. 1982, \aap, 114, 245

\bibitem[{{Udry} \& {Santos}(2007)}]{udry2007}
{Udry}, S. \& {Santos}, N.~C. 2007, \araa, 45, 397

\bibitem[{{Umebayashi} \& {Nakano}(2009)}]{umebayashi2009}
{Umebayashi}, T. \& {Nakano}, T. 2009, \apj, 690, 69

\bibitem[{{Vasyunin} {et~al.}(2008){Vasyunin}, {Semenov}, {Henning}, {Wakelam},
  {Herbst}, \& {Sobolev}}]{vasyunin2008}
{Vasyunin}, A.~I., {Semenov}, D., {Henning}, T., {et~al.} 2008, \apj, 672, 629

\bibitem[{{Visser} {et~al.}(2015){Visser}, {Bergin}, \&
  {J{\o}rgensen}}]{visser2015}
{Visser}, R., {Bergin}, E.~A., \& {J{\o}rgensen}, J.~K. 2015, \aap, 577, A102

\bibitem[{{Visser} {et~al.}(2009){Visser}, {van Dishoeck}, {Doty}, \&
  {Dullemond}}]{visser2009}
{Visser}, R., {van Dishoeck}, E.~F., {Doty}, S.~D., \& {Dullemond}, C.~P. 2009,
  \aap, 495, 881

\bibitem[{{Walsh} {et~al.}(2010){Walsh}, {Millar}, \& {Nomura}}]{walsh2010}
{Walsh}, C., {Millar}, T.~J., \& {Nomura}, H. 2010, \apj, 722, 1607

\bibitem[{{Walsh} {et~al.}(2012){Walsh}, {Nomura}, {Millar}, \&
  {Aikawa}}]{walsh2012}
{Walsh}, C., {Nomura}, H., {Millar}, T.~J., \& {Aikawa}, Y. 2012, \apj, 747,
  114

\bibitem[{{Walsh} {et~al.}(2015){Walsh}, {Nomura}, \& {van Dishoeck}}]{walsh15}
{Walsh}, C., {Nomura}, H., \& {van Dishoeck}, E. 2015, \aap, 582, A88

\bibitem[{{Weidenschilling}(1977)}]{weidenschilling1977}
{Weidenschilling}, S.~J. 1977, \apss, 51, 153

\bibitem[{{Willacy}(2007)}]{willacy2007}
{Willacy}, K. 2007, \apj, 660, 441

\bibitem[{{Willacy} {et~al.}(1998){Willacy}, {Klahr}, {Millar}, \&
  {Henning}}]{willacy1998}
{Willacy}, K., {Klahr}, H.~H., {Millar}, T.~J., \& {Henning}, T. 1998, \aap,
  338, 995

\bibitem[{{Williams} \& {Cieza}(2011)}]{williams11}
{Williams}, J.~P. \& {Cieza}, L.~A. 2011, \araa, 49, 67

\bibitem[{{Wirstr{\"o}m} {et~al.}(2016){Wirstr{\"o}m}, {Lerner},
  {K{\"a}llstr{\"o}m}, {Levinsson}, {Olivefors}, \& {Tegehall}}]{wirstrom2016}
{Wirstr{\"o}m}, E.~S., {Lerner}, M.~S., {K{\"a}llstr{\"o}m}, P., {et~al.} 2016,
  \aap, submitted

\bibitem[{{Woitke} {et~al.}(2009){Woitke}, {Kamp}, \& {Thi}}]{woitke2009}
{Woitke}, P., {Kamp}, I., \& {Thi}, W.-F. 2009, \aap, 501, 383

\bibitem[{{Zhang} {et~al.}(2015){Zhang}, {Blake}, \& {Bergin}}]{zhang2015}
{Zhang}, K., {Blake}, G.~A., \& {Bergin}, E.~A. 2015, \apjl, 806, L7

\end{thebibliography}

\begin{appendix} 
\section{Timescales}
\label{timescale-app}
\begin{figure*}
\subfigure{\includegraphics[width=0.5\textwidth]{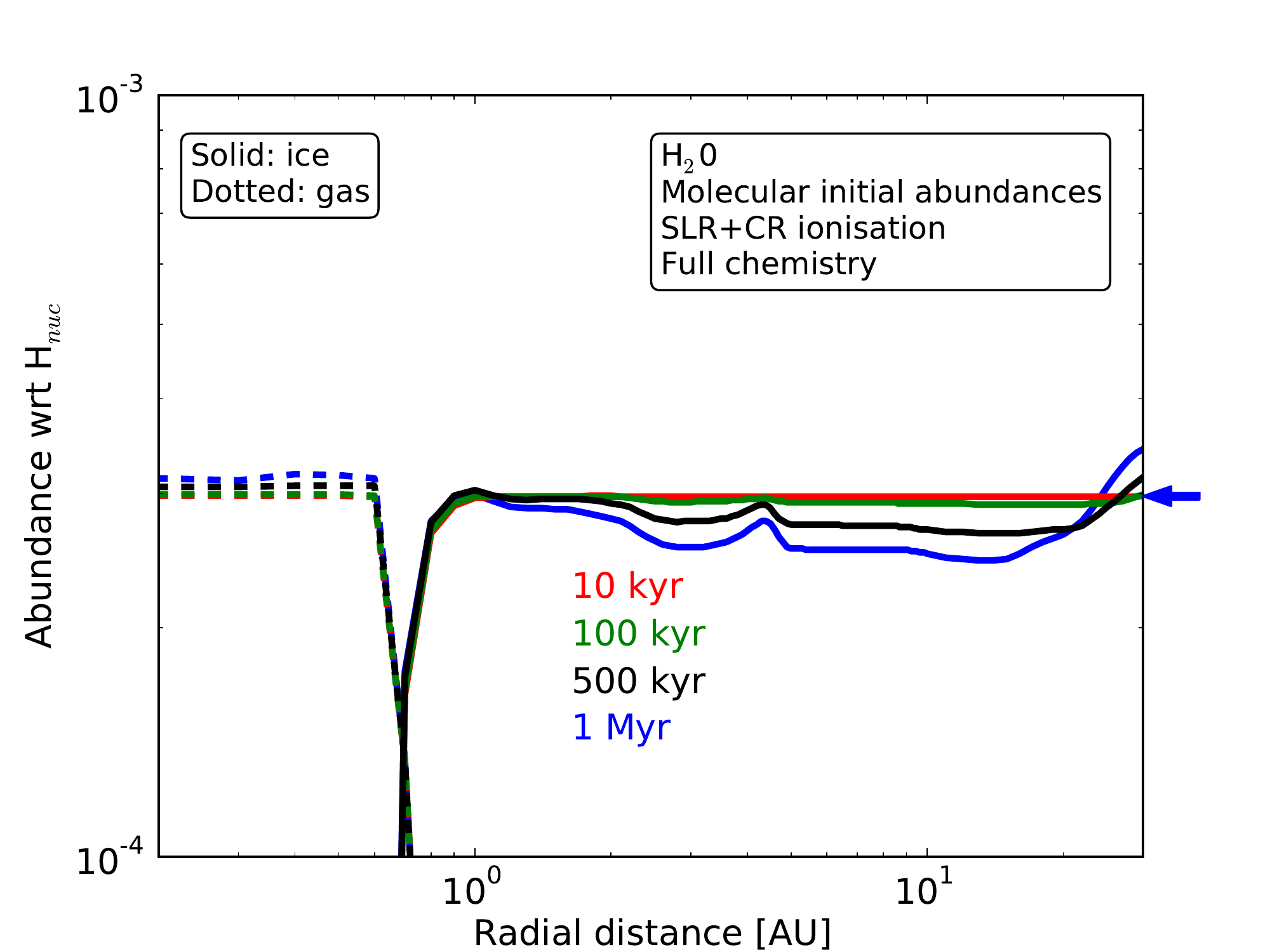}\label{h2oevol}}
\subfigure{\includegraphics[width=0.5\textwidth]{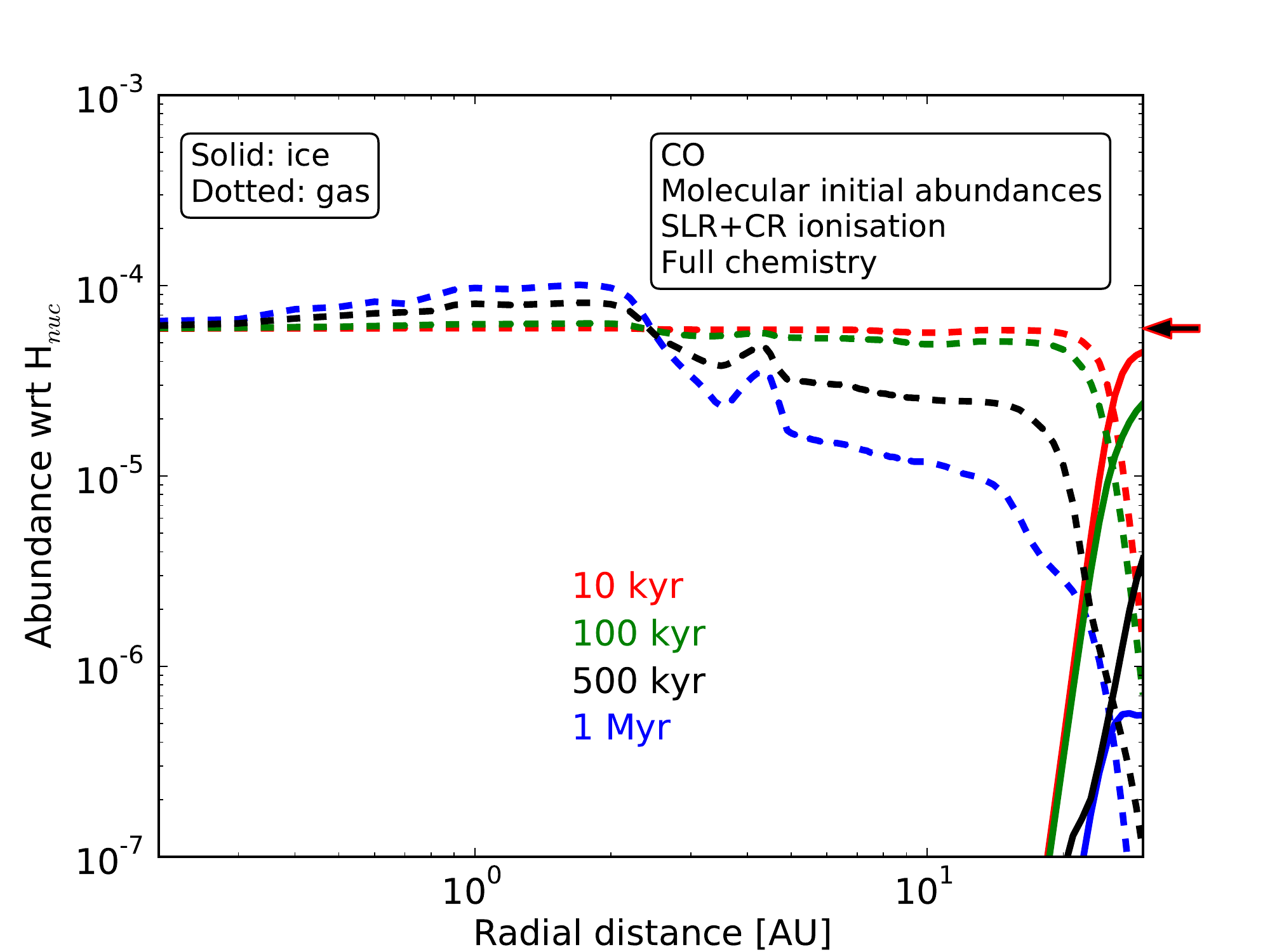}\label{coevol}}\\
\subfigure{\includegraphics[width=0.5\textwidth]{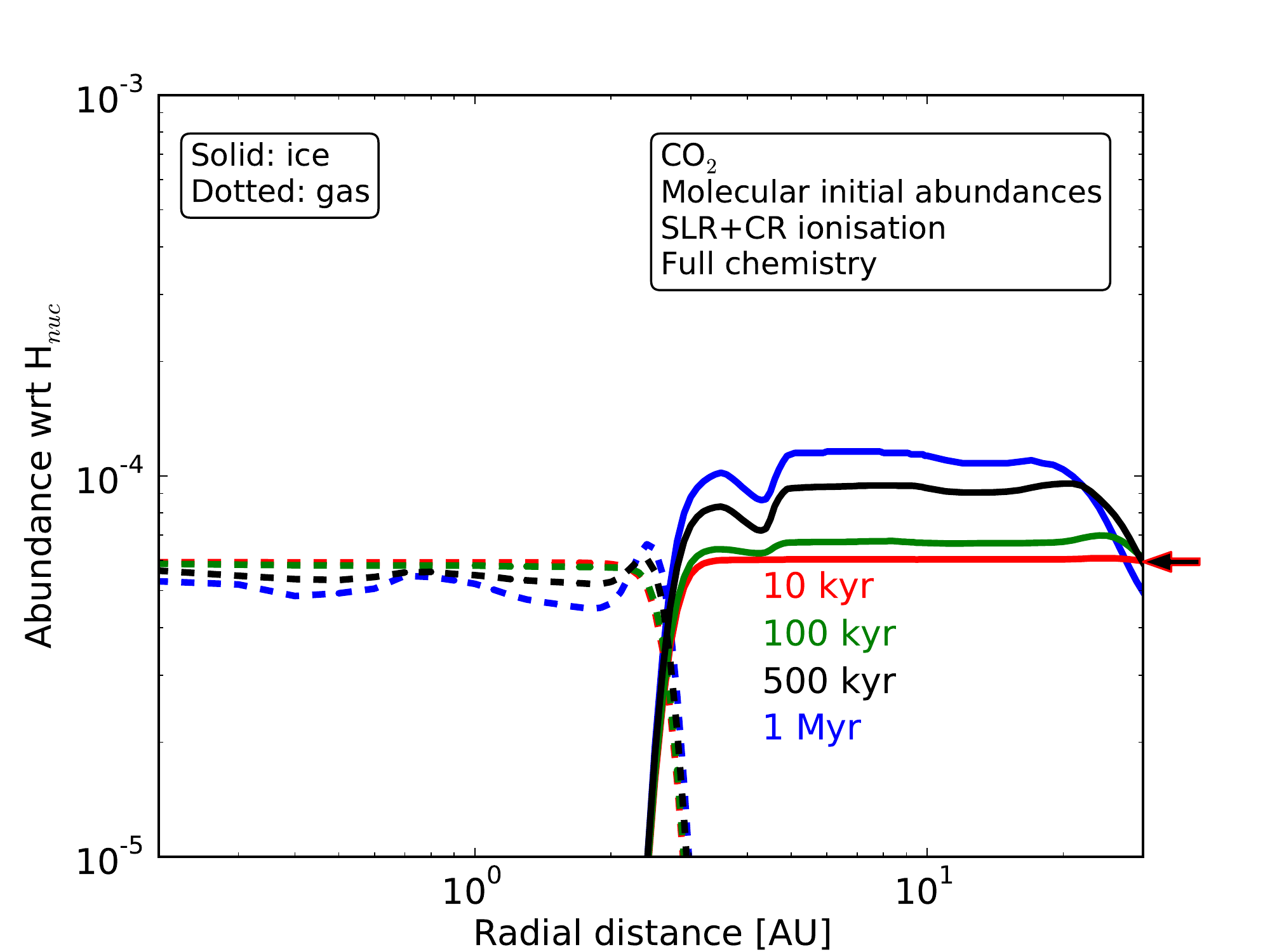}\label{co2evol}}
\subfigure{\includegraphics[width=0.5\textwidth]{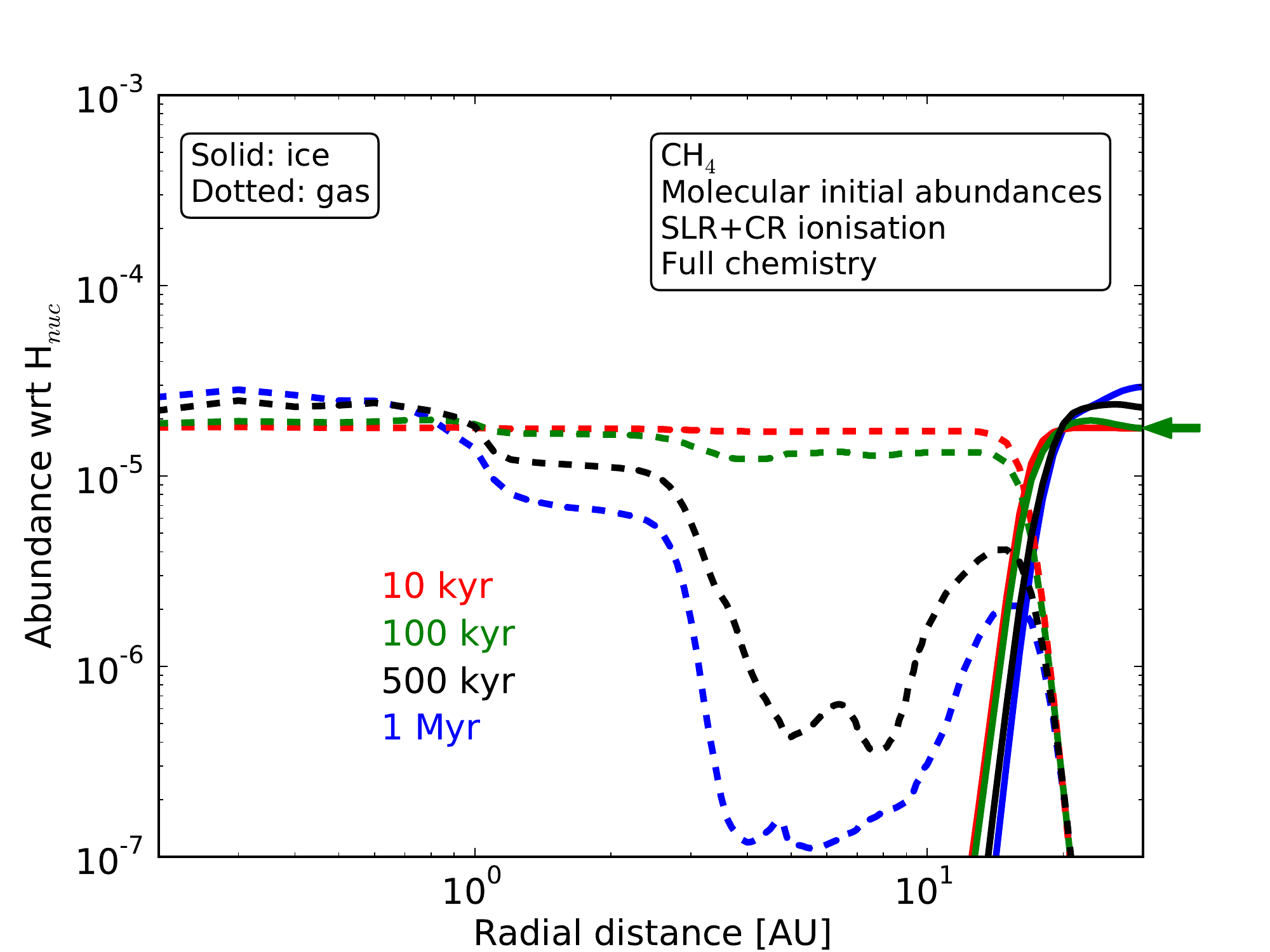}\label{ch4evol}}
\caption{Abundances as function of radial distance $R$ of the four key volatiles, taken at four different evolutionary stages. Note the different y-axes ranges. Arrows to the right of each panel indicate the initial abundance level in the case of the inheritance-scenario.}
\label{evolutions}
\end{figure*}

Fig. \ref{evolutions} shows the abundances of the four key volatile, taken at four different evolutionary times. Note the different y-axes ranges. It is seen that significant changes occur between 100 and 500 kyr, that is after a few times $10^{5}$ yrs. This goes for all the four species, gas as well as ice, out to about 25 AU.

It is interesting to see that \ce{H2O} and \ce{CO2} are respectively produced and destroyed in the gas phase in the inner disk, and vice versa in the outer disk. For CO and \ce{CH4} there are turn-over radii in the inner disk, inside of which they are both being produced, and outside of which they are both being destroyed. These turn-over radii correspond to the iceline positions of \ce{H2O} and \ce{CO2} for the case of \ce{CH4} and CO respectively. This nicely illustrates the dependance on gas phase \ce{H2O} and \ce{CO2} for the production/destruction of CO and \ce{CH4} in the gas phase.

\end{appendix}

\end{document}